\documentclass[12pt,a4paper]{article}
\pdfoutput=1
\usepackage{amsmath,amsfonts,amssymb,amsthm,bbm,bm}
\usepackage{pdfsync}
\usepackage{mathtools}
\usepackage{graphicx}
\usepackage[latin1]{inputenc}
\usepackage{hyperref}
\usepackage[all]{xy}
\usepackage{stmaryrd}
\usepackage{rotating}
\usepackage{color}  
\usepackage{slashed}
\usepackage[dvipsnames]{xcolor}
\usepackage{comment}
\usepackage{cite}
\usepackage{stackrel}

\newcommand{\bc}{\begin{center}}
\newcommand{\ec}{\end{center}}

\newcommand{\cL}{{\mathcal L}}

\newcommand{\be}{\begin{align}}
\newcommand{\ee}{\end{align}}
\newcommand{\bea}{\begin{eqnarray}}
\newcommand{\eea}{\end{eqnarray}}
\newcommand{\bs}{\begin{subequations}}
\newcommand{\es}{\end{subequations}}
\newcommand{\nn}{\nonumber}

\newcommand{\tl}{\widetilde}
\def\p{\partial}

\newcommand{\pd}{\partial}

\def\de{\delta}
\def\eps{\epsilon}

\def\th{\theta}

\allowdisplaybreaks

\title{{\bf \!\!\!\!\!\!\!\! Metric reconstruction from celestial multipoles\!\!\!\!\!\!\!\!}}

\date{}

\begin{document}

\maketitle

\vspace{-1.5cm}

\centerline{\large{\bf Geoffrey Comp\`{e}re,$^{M_{L,k}}$\footnote{geoffrey.compere@ulb.be}, 
Roberto Oliveri,$^{S_{L,k}}$\footnote{roberto.oliveri@obspm.fr},
Ali Seraj$^{\,M_{L,k}}$\footnote{ali.seraj@ulb.be}
}}\vspace{6pt}

\medskip
\centerline{\textit{{}$^{M_{L,k}}$ Universit\'{e} Libre de Bruxelles, Centre for Gravitational Waves, }}
\centerline{\textit{International Solvay Institutes, CP 231, B-1050 Brussels, Belgium}}

\medskip
\centerline{\textit{{}$^{S_{L,k}}$ LUTH, Laboratoire Univers et Th\'eories, Observatoire de Paris}}
\centerline{\textit{CNRS, Universit\'e PSL, Universit\'e Paris Cit\'e,}}
\centerline{\textit{5 place Jules Janssen, 92190 Meudon, France}}
\vspace{1cm}

\begin{abstract}
\noindent 
 The most general vacuum solution to Einstein's field equations with no incoming radiation can be constructed perturbatively from two infinite sets of canonical multipole moments, which are found to be mapped into each other under gravitational electric-magnetic duality at the non-linear level. We demonstrate that in non-radiative regions such spacetimes are completely characterized by a set of conserved celestial charges that consist of the Geroch-Hansen multipole moments, the generalized BMS charges and additional celestial multipoles accounting for subleading memory effects.
 Transitions among non-radiative regions, induced by radiative processes, are therefore labelled by celestial charges, which are identified in terms of canonical multipole moments of the linearized gravitational field. The dictionary between celestial charges and canonical multipole moments allows to holographically reconstruct the metric in de Donder, Newman-Unti or Bondi gauge outside of sources.
\end{abstract}


\setcounter{footnote}{0}

\newpage
\tableofcontents

\section{Introduction and outline}
\label{sec:intro}

The $w_{1+\infty}$ algebra and its loop generalization $Lw_{1+\infty}$ have been identified recently in the structure of asymptotically flat spacetimes using techniques from celestial holography, as a consequence of the infinite tower of tree-level soft graviton theorems  \cite{Strominger:2021lvk,Strominger:2021mtt,Guevara:2021abz,Freidel:2021ytz}. Classically, such asymptotic symmetries act on the charges that obey local flux-balance laws at null infinity and which are conserved at spatial and timelike infinity. A universal formulation for these conserved quantities in terms of higher spin-weight operators was found and shown to reproduce the classical soft theorems and the $Lw_{1+\infty}$ algebra structure at least in the linear theory \cite{Freidel:2021ytz}. This uncovers an infinite-dimensional symmetry structure that supplements the previously found BMS, dual BMS and generalized BMS asymptotic symmetry algebras  \cite{Bondi:1960jsa,1962RSPSA.269...21B,1962PhRv..128.2851S,10.2307/2415610,1962RSPSA.270..103S,Barnich:2009se,Barnich:2010eb,Barnich:2011mi,Kapec:2014opa,Campiglia:2014yka,Campiglia:2020qvc,Compere:2018ylh,Godazgar:2018vmm,Godazgar:2018qpq,Godazgar:2018dvh,Oliveri:2020xls,Freidel:2021fxf,Freidel:2021qpz,Freidel:2021dfs}.
In this paper, we will elaborate on the physical meaning of the $Lw_{1+\infty}$ conserved charges, and their relationship to the multipole expansions of the gravitational field close to null and spatial infinity. The non-linear expansion of the metric close to null infinity has long been known in the Bondi-Sachs expansion  \cite{1962PhRv..128.2851S,10.2307/2415610,Tamburino:1966zz,1985FoPh...15..605W,Barnich:2010eb}, but, only recently, all the local flux-balance laws have been explicitly written down in a form amenable to analysis from the point of view of infinite towers of soft theorems and memory effects \cite{Grant:2021hga}. 
All Bondi aspects, defined at increasing orders $1/r^{n+1}$ in the radial expansion around null infinity, \emph{i.e.} the mass aspect ($n=0$) in $g_{uu}$, the angular momentum aspect ($n=1$) in $r^{-1} g_{ua}$ and the infinite tower of 2-dimensional symmetric trace-free (STF) Bondi aspects ($n \geq 2$) in the expansion of the spherical components $r^{-2} g_{ab}$, obey evolution equations whose fluxes contain non-linear interactions of the Bondi shear, news and aspects at most of order ${\text{max}\{2,n+1\}}$. 
These evolution equations can be written as an equality between linear terms in the Bondi news, sometimes called ``soft'' or ``memory'' terms, and the sum of retarded time derivatives of Bondi aspects and non-linear flux terms  which provide the source of ordinary and null memories, respectively.
In order to express an arbitrary $n \geq 0$ Bondi aspect in terms of the source causing gravitational radiation, further information is required beyond the Bondi-Sachs expansion. One approach consists in combining the Bondi-Sachs expansion with the multipolar post-Minkowskian expansion in de Donder/harmonic gauge \cite{Blanchet:1985sp,Blanchet:1986dk,Blanchet:1987wq,Blanchet:1992br}.
The dictionary between the so-called canonical moments $M_L(u)$ and $S_L(u)$, that parametrize a generic post-Minkowskian metric without incoming radiation, and the Bondi aspects have been derived so far at linear order and for quadratic tail and memory terms \cite{Blanchet:2020ngx}. 
In this work, we first define the complete set of (real-valued) celestial charges that label a generic vacuum asymptotically flat spacetime around null infinity with no incoming radiation. We further re-express the (real-valued) celestial charges in terms of the complex basis of helicity $s$ charges as defined in \cite{Freidel:2021dfs,Freidel:2021ytz} at the linear level. The  celestial charges 
\begin{equation} \label{eq:1}
\mathcal P_L(u), \; \mathcal P_L^-(u), \; \mathcal K_L(u),\; \mathcal J_L(u), \;\mathcal Q^\pm_{n,L}(u),\; n \geq 2,  
\end{equation}
consist of the ``supermomenta'' $\mathcal P_L$, the ``dual supermomenta'' $\mathcal P_L^-$, the ``super-center-of-mass'' $\mathcal K_L$, the ``super-angular momenta'' $\mathcal J_L$, and the ``$n \geq 2$ celestial charges'' $\mathcal Q^\pm_{n,L}$ for any $n \geq 2$. 
The non-linear metric in Bondi gauge is uniquely determined, up to a residual BMS transformation, in terms of the past history of the news tensor as well as the celestial charges at early retarded times \cite{1962PhRv..128.2851S,10.2307/2415610,Tamburino:1966zz,1985FoPh...15..605W,Barnich:2010eb,Grant:2021hga,Freidel:2021dfs}. We distinguish two qualitatively distinct classes of celestial charges depending on whether or not they contain a linear term in the news (\emph{i.e.} the ``memory'' term) in their associated flux-balance law. Discarding NUT charges, the ``memory-less charges'' are the ten Poincar\'e charges and the $n \geq 3$ Newman-Penrose charges defined as the celestial charges $\mathcal Q^\pm_{n,L}(u)$ with $2\leq \ell \leq n-1$. The complementary ``memory-full charges'' are the complementary set of celestial charges. The motivation for this distinction is that the corresponding flux-balance laws have different physical implications. The memory-full flux-balance laws can be used to determine the memory effects due to the radiation process \cite{Bieri:2013ada,Pasterski:2015tva,Flanagan:2015pxa,Nichols:2018qac,Mitman:2020pbt,Mitman:2021xkq}, while the memory-less flux-balance laws characterize radiation-reaction effects, \emph{i.e.} they determine how the source reacts to the emitted radiation \cite{1964PhRv..136.1224P,PhysRev.131.435,Damour:1983tz,1983MNRAS.203.1049F,Blanchet:2013haa,Buonanno:2014aza,Blanchet:2018yqa,Compere:2019gft}. 
Second, we describe the action of linear gravitational $SO(2)$ electric-magnetic duality \cite{Henneaux:2004jw} in terms of canonical multipole moments. We find that the even parity $M_L^+(u) \equiv M_L(u)$ and odd parity $M_L^-(u)\equiv \frac{2 \ell}{\ell+1} S_L(u)$ canonical multipoles are rotated into each other at the linear level. Moreover, since a non-linear metric with no incoming radiation can be perturbatively constructed out of the canonical multipole moments, we find that any non-linear extension of the linear duality defined above induces an automorphism on the space of solutions. Imposing that the symplectic structure at null infinity and the celestial charges at past null infinity are  preserved under duality rotations fixes unique non-linear corrections and thus defines the gravitational electric-magnetic duality at the non-linear level. This provides a non-linear perturbative completion of the duality described in \cite{Henneaux:2004jw} at the level of the non-linear solution space.
The third and main result of this article is that in non-radiative regions, the celestial charges split into two qualitatively distinct sets of conserved charges: (i) the non-radiative multipole charges $M^\pm_{L,k}$ with $0 \leq k \leq \ell$ that entirely parameterize non-radiative spacetimes without incoming radiation, and (ii) the $n \geq 3$ Newman-Penrose charges.
The non-radiative multipole charges are in one-to-one correspondence with the Geroch-Hansen multipole moments, the generalized BMS charges (including the Poincar\'e charges) together with the $n \geq 2$ celestial  charges $\mathcal Q^\pm_{n,L}$ with $n \leq \ell$. As we will argue in the main text, the non-radiative multipole charges completely determine an arbitrary non-radiative spacetime.  Radiative processes can be therefore thought of as inducing a transition from an initial to a final state both labelled by a set of celestial charges $M^\pm_{L,k}$ with $0 \leq k \leq \ell$. The towers of conserved non-radiative multipole charges are summarized in Tables \ref{tab:1} and \ref{tab:2}. They can be decomposed as follows: the charges $M^+_{L,k}$ for $\ell=0,1$, as well as $M^-_{L,0}$ for $\ell=1$ are the Poincar\'e charges, while $M^\pm_{L,0}$ for $\ell \geq 2$ are respectively the stationary mass and spin multipole moments defined by Geroch \cite{Geroch:1970cd} and Hansen \cite{Hansen:1974zz}, up to a normalization constant. For the highest value $k = \ell$, the charges $M^+_{L,\ell}$ are the supermomenta proportional to the constant part of the shear, \emph{i.e.} the displacement memory field, while $M^-_{L,\ell}$ for $\ell \geq 1$ are the dual supermomenta \cite{Godazgar:2018vmm,Godazgar:2018qpq,Godazgar:2018dvh} that are usually set to vanish for standard boundary conditions.  The intermediate charges for $1 \leq k \leq \ell-1$ are super-Lorentz \cite{Barnich:2009se,Barnich:2010eb,Barnich:2011mi,Kapec:2014opa,Campiglia:2014yka,Campiglia:2020qvc,Compere:2018ylh} and $n \geq 2$ celestial charges \cite{Freidel:2021qpz,Freidel:2021dfs} related to subleading memory effects \cite{Pasterski:2015tva,Nichols:2017rqr,Nichols:2018qac,Flanagan:2018yzh,Compere:2019odm,Grant:2021hga,Seraj:2021qja,Seraj:2021rxd,Seraj:2022qyt}. 
The $n \geq 3$ Newman-Penrose charges are the complementary subset of celestial charges, namely $\mathcal Q^\pm_{n,L}$ with $2 \leq \ell \leq n-1$. They identically vanish in the linearized theory in non-radiative regions.  In linearized gravity, canonical multipole moments therefore suffice to characterize a generic non-radiative spacetime, as is well-known, see \emph{e.g.} \cite{Blanchet:2013haa}. In non-radiative regions, the $n \geq 3$ Newman-Penrose charges are non-vanishing once non-linear interactions are included, but, by completeness, they are functionals of the non-radiative multipole charges. The distinguished example of such non-linear conserved charges are the 10  $n=3$ Newman-Penrose charges \cite{1968RSPSA.305..175N} $\mathcal Q^\pm_{3,ij}$ corresponding the lowest order $n=3$, $\ell=2$, which have indeed been proven to be non-linear combinations of non-radiative multipole moments \cite{10.2307/2415610}. Interesting, all  Newman-Penrose charges have the property that their associated flux-balance laws are memory-less: they admit no ``soft'' term, on the same footing as the 10 Poincar\'e flux-balance laws. 
The identification of the complete set of conserved charges in non-radiative regions such as at spatial and timelike infinity provides a new interesting perspective in the programme of reconstructing an asymptotically flat metric from holographic data.
The method that we employed, the post-Minkowskian formalism formulated in de Donder gauge combined with a Bondi expansion is perfectly suited for that purpose since it combines the reconstruction of the bulk metric in de Donder gauge with the identification of the holographic data in Bondi gauge. Much work remains to be performed to write a full non-linear metric to order $G^2$ and $G^3$ outside of sources in terms of celestial charges. 
Multipole moments can be defined in a gauge invariant way at least at spatial infinity \cite{Geroch:1970cd,Hansen:1974zz}. The dictionary that we derived between the multipole moments and the complex charges forming the $Lw_{1+\infty}$ algebra (at the linear level so far) brings promising perspectives in the gauge invariant quantization of the gravitational field with asymptotically flat boundary conditions. 
\paragraph{Notation.} We use $a,b,c,d$ to refer to coordinates on the sphere, and $i,j,\cdots ,n$ to denote Cartesian indices. We implement the multi-index notation of Ref. \cite{Blanchet:2020ngx},  where $L\equiv i_1 \dots i_\ell$, with $i_k = \{ x,y,z\}$ being Cartesian indices. Multi-index tensors used here are symmetric trace-free (STF), unless otherwise mentioned. The STF part of a tensor is denoted by angle brackets $\langle\cdots \rangle$ around
its indices. Embedding the celestial sphere in $\mathbb{R}^3$, one defines $n_i\equiv x_i/r$. The STF harmonics $\hat n_L\equiv n_{\langle i_1}\cdots n_{i_\ell\rangle }$ represent the rotation group. A generic function on the sphere can be decomposed as $T(\th,\phi)=T_L \hat{n}_L$. A natural basis for the sphere embedded in $\mathbb{R}^3$ is $e_a{}^i=\frac{\pd n^i}{\pd \th^a}$ that can be used to project a tensor onto the sphere. We define the dual of forms and vectors on the sphere as  $\widetilde V_a = \epsilon_{a}^{\; \; b}V_b$, $\widetilde V^a = \epsilon^{a}_{\;\; b}V^b$. For any STF tensor $T^{ab}$ on the sphere, we define its dual STF tensor as $\widetilde T^{ab}\equiv \epsilon^{ac}T_{c}^{\;\;b}=\epsilon^{b}_{\;\; c}T^{ac}$.  Similarly, the dual of a higher rank STF tensor is defined from the dualization of the first tensorial index.

\section{Local flux-balance laws and conserved charges}
\label{sec:fbl}

In a Bondi coordinate system $(u,r,\th^a)$ and after expanding the metric in integer powers of $1/r$, Einstein's equations reduce to a set of algebraic constraints except for a countable infinite set of local flux-balance equations on future null infinity $\mathcal I^+$. Such local flux-balance equations, once integrated over the range $-\infty < u < \infty$, give rise to a corresponding countable infinite set of global flux-balance laws on $\mathcal I^+$, labelled by an integer $n \geq 0$, which upon quantization leads to the sub$^n$-leading  soft graviton theorems \cite{Weinberg:1965nx,Cachazo:2014fwa,Hamada:2018vrw}. 

The spherical metric $g_{ab}$ in Bondi gauge takes the form \cite{Grant:2021hga}
\begin{equation}
g_{ab}=r^2 \sqrt{1+\frac{\mathcal C_{cd}\mathcal C^{cd}}{2r^2}} \gamma_{ab}+r\,  \mathcal C_{ab}, \qquad \mathcal C_{ab} \equiv C_{ab}+\sum_{n=2}^\infty r^{-n} \stackrel[(n)]{}{E_{ab}},
\end{equation}
where $\mathcal C_{ab}$ is STF with respect to the unit sphere metric $\gamma_{ab}$. The news tensor is defined as $N_{ab}=\partial_u C_{ab}$. Our conventions for the Bondi mass aspect $m$ and angular momentum aspect $N_a$ can be read from the metric components 
\begin{align}
   g_{uu} &=-1+\frac{2m}{r} + \mathcal{O}(r^{-2}), \\
	g_{ua}&=\frac{1}{2}D^b C_{ab}+\frac{2}{3r}\left( N_a -\frac{3}{32}D_a (C_{bc}C^{bc})\right)+ \mathcal{O}(r^{-2}),
\end{align}
and agree with \cite{Flanagan:2015pxa,Grant:2021hga}

\subsection{Local flux-balance laws and Bondi aspects}

The local flux-balance laws take the form~\footnote{The local flux-balance laws are invariant under the field redefinition $Q_{\dots} \rightarrow Q_{\dots}+\Delta Q_{\dots}$, $\mathcal F_{\dots} \rightarrow \mathcal F_{\dots}+\partial_u \Delta \mathcal Q_{\dots}$ for each charge aspect $Q_{\dots}=m,\, \mathcal N_a,\dots$ and corresponding flux $\mathcal F,\, \mathcal F_a,\dots$. This ambiguity is partially fixed by the requirement that the fluxes are at least linear in the news. One should then add more terms at least linear in the shear to the charges in order to ensure that the charges are invariant under a change of foliation. Such a program has been completed only for $n=0$ and $n=1$ and the corresponding expressions are given here. The expressions for $n \geq 2$ instead follow \cite{Grant:2021hga}.} 
\begin{subequations}\label{fb}
\begin{eqnarray}
n=0&: \quad & \frac{1}{4}D_b D_c N^{bc} = -\mathcal F(u)+ \partial_u m , \label{FBL0} \\
n=1&: \quad &-\frac{u}{2}D_c D_{\langle a}D_{b\rangle} {N}^{bc} = -\mathcal F_a(u)+ \partial_u \mathcal N_a, \label{FBL1}\\
n=2&: \quad &\frac{u^2}{12}\text{STF}_{ab}[D_a D_c D_{\langle b}D_{d\rangle}  {N}^{cd}] = - \stackrel[(2)]{}{\mathcal F_{ab}}(u)+\partial_u  \stackrel[(2)]{}{\mathcal E_{ab}}, \label{FBL2}\\ 
n\geq 3&:  \quad &\frac{(-u)^n}{6\,  n!}\mathcal D_{n-3}\cdots \mathcal D_0\, \text{STF}_{ab}[D_a D_c D_{\langle b}D_{d\rangle}  {N}^{cd} ]= - \stackrel[(n)]{}{\mathcal F_{ab}}(u)+\partial_u  \stackrel[(n)]{}{\mathcal E_{ab}}. \label{FBL3}
\end{eqnarray}
\end{subequations}

The differential operator $\mathcal D_n$ on the sphere for any $n \geq 0$ integer is defined as \cite{Grant:2021hga}
\begin{equation}\label{D_n def}
\mathcal D_n \equiv -\frac{n+2}{2(n+1)(n+4)} \left(\Delta +n^2+5n+2\right).
\end{equation}
In a multipolar decomposition, the operator $\mathcal D_n$ annihilates only the $\ell = n+2$ spin 2 tensor harmonics. Explicitly, any STF tensor $T_{ab}$ can be decomposed as
\begin{equation}\label{decompT}
T_{ab}=-2 D_{\langle a} D_{b \rangle} T^+ +2 \epsilon_{c (a} D_{b)} D^c T^-. 
\end{equation}
The harmonic modes $T^\pm=T^\pm_L \hat{n}_L$ constitute the kernel of $\mathcal D_{\ell-2}$. Therefore, $\mathcal D_{n-3}\cdots \mathcal D_0 T_{ab}$ annihilates all harmonic modes of $T^\pm$ of rank $\ell=2,\cdots n-1$. 

We also define the tensor that includes the Bondi mass aspect and dual Bondi mass aspect \cite{Godazgar:2018qpq,Godazgar:2018dvh,Godazgar:2019dkh}
\begin{equation}
m_{ab} \equiv m \gamma_{ab} + \frac{1}{2}D_{[a}D^c C_{b]c}= 
m \gamma_{ab} +m^-\epsilon_{ab}, \qquad m^-\equiv\frac{1}{4}D_c D_d \tl{C}^{cd},
\end{equation}
where $\tl{C}^{ab}\equiv \eps^{ac}C_c{}^b$ is the dual shear. Here, $\eps_{ab}$ defines a complex structure as $\eps_{ac}\eps^{cb}=-\de_a{}^b$, and therefore there is an isomorphism between $m \gamma_{ab} +m^-\epsilon_{ab}$ and the complex number $m+i m^-$.

The finite energy fluxes, sometimes called the ``hard'' terms, which are defined to vanish when the news vanishes, are given by\footnote{Note that one can show 
$\mathcal F_a=\frac{1}{4}\mathcal H_a(N_{bc},C_{bc})-u D_a \mathcal F$ where $\mathcal H_a(N_{bc},C_{bc})$ is the hard super-Lorentz operator \cite{Campiglia:2016hvg,Compere:2019gft} after using the property $C_{ac}D_b N^{bc}=C^{bc}D_aN_{bc}-C^{bc}D_bN_{ac}$ valid for any pair of STF tensors on the 2-sphere.}
\begin{align}
\mathcal F &\equiv   -\frac{1}{8}N_{ab}N^{ab}, \nonumber \\
\mathcal F_a &\equiv  \frac{1}{4}N^{bc}D_bC_{ca}+\frac{1}{2}C_{ab}D_c N^{bc}- \frac{1}{4}N_{ab}D_c C^{bc} -\frac{1}{8}\partial_a (C_{bc}N^{bc})-u D_a \mathcal F, \\ 
\stackrel[(2)]{}{\mathcal F_{ab}} &\equiv  \frac{1}{4}N_{cd}C^{cd}C_{ab}- \frac{u}{2}\partial_u (C_{a}^{\; c}m_{bc})-uD_{\langle a}\left(\frac{1}{12} N^{cd}D_{\vert d}C_{c \vert b \rangle}+C_{b\rangle c}D_d N^{cd}\right) +\frac{u^2}{6}D_{\langle a} D_{b\rangle}\mathcal F , \nonumber
\end{align}
while $\stackrel[(n)]{}{\mathcal F_{ab}} \,=\, \mathcal D_{n-2}\stackrel[(n-1)]{}{\mathcal E_{ab}}+\dots$, where dots indicate non-linear terms.

The Bondi aspects appearing in the flux-balance laws are: for $n=0$ the Bondi mass aspect $m$; for $n=1$ the ``dressed angular momentum aspect'' \cite{Compere:2018ylh,Compere:2019gft}
\begin{equation}
 \mathcal N_a \equiv N_a-\frac{1}{4}C_{ab}D_c C^{bc}-\frac{1}{16} \partial_a (C_{bc}C^{bc})- u D^b m_{ab};
\end{equation}
for $n=2$ the ``dressed $n=2$ Bondi aspect''
\begin{equation}
\stackrel[(2)]{}{\mathcal E_{ab}}
\, = \, \stackrel[(2)]{}{E_{ab}}-\frac{u}{2}C_{(a}^{\;\; c}m_{b)c} -\frac{u}{3}D_{\langle a}  \mathcal N_{b\rangle} -\frac{u^2}{6}D_{\langle a}D^cm_{b \rangle c},
\end{equation}
and for $n \geq 3$ the ``dressed $n \geq 3$ Bondi aspects''
\begin{align}\label{Eabn}
\stackrel[(n)]{}{\mathcal E_{ab}} \, = \, \stackrel[(n)]{}{E_{ab}} - u \stackrel[(n-1,-1)]{}{\mathcal G_{ab}} &+\sum_{p=1}^{n-2} \frac{(-u)^{p}}{p!}\mathcal D_{n-3}\cdots \mathcal D_{n-2-p} \left( \stackrel[(n-p)]{}{E_{ab}} -\frac{u}{p+1}\stackrel[(n-1,p-1)]{}{\mathcal G_{ab}} \right)\nonumber\\
&+  \frac{(-u)^{n-1}}{3(n-1)!} \mathcal D_{n-3} \cdots \mathcal D_0 D_{\langle a} \left( N_{b \rangle} -\frac{u}{n}D^cm_{b\rangle c} \right) ,
\end{align}
where $\stackrel[(n-1,p)]{}{\mathcal G_{ab}}$ are non-linear terms defined in \cite{Grant:2021hga}.

Corresponding to $n=0,1$, following the conventions of \cite{Compere:2019gft}, the generalized BMS charges are defined as 
\begin{align}\label{BMS01}
 \mathcal P_L = \oint_S m \, \hat n_L ,\qquad  -\mathcal J_L=\frac{1}{2} \oint_S \epsilon^{ab}\partial_b \hat n_L \, \mathcal N_a,\qquad   \mathcal K_L=\frac{1}{2}\oint_S \partial^a \hat n_L\,  \mathcal N_a, 
\end{align}
while the dual supermomenta are defined as \cite{Godazgar:2018qpq,Godazgar:2018dvh,Godazgar:2019dkh} 
\begin{align}\label{BMS012}
\mathcal P^-_L = \oint_S  m^- \, \hat n_L = \frac{1}{2}\oint_S m_{ab}\epsilon^{ab} \, \hat n_L.  
\end{align}
The 10 Poincar\'e charges are $\mathcal E \equiv \mathcal P_{\emptyset}$, $\mathcal P_i$, $\mathcal J_i$, $\mathcal K_i$.

Using Eq.~\eqref{decompT}, for each $n \geq 2$, we decompose the field $\stackrel[(n)]{}{\mathcal E_{ab}}$ into its parity even and parity odd $\ell \geq 2$ harmonic components, which allows to define the parity even and parity odd ``$n \geq 2$ celestial charges'' 
\begin{eqnarray}
\mathcal Q^+_{n,L}(u) \equiv   \oint_S \stackrel[(n)]{}{\mathcal E^{ab}} D_{a} D_{b} \hat n_L  ,\qquad
\mathcal Q^-_{n,L}(u)  \equiv \oint_S \stackrel[(n)]{}{\mathcal E^{ab}} \epsilon_{ac} D_{b} D^c \hat n_L  .\label{higherspincharges}
\end{eqnarray}The $n \geq 2$ celestial charges $\mathcal Q^\pm_{n,L}$ vanish by construction for $\ell=0,1$.  We will not attempt to associate the charges \eqref{higherspincharges} with asymptotic symmetries and we will therefore restrain from calling them Noether/Hamiltonian/canonical charges. They are yet surface charges in the sense that they are defined from an integral over the sphere of dressed Bondi aspects exactly as the generalized BMS charges. We note however that $\mathcal Q_{\ell-1,L}^+$ is the momentum multipole moment in the language of \cite{Compere:2017wrj} which are defined as Noether charges associated with momentum multipole symmetries; see Eq.~(4.10) of \cite{Compere:2017wrj}.

We can parameterize the entire set of celestial charges including the BMS and dual BMS charges by introducing the notation 
\begin{eqnarray}\label{defQlow}
\mathcal Q^+_{0,L}(u) \equiv \mathcal P_L(u), \quad \mathcal Q^-_{0,L}(u) \equiv \mathcal P^-_L(u), \quad \mathcal Q^+_{1,L}(u) \equiv \mathcal K_L(u),  \quad \mathcal Q^-_{1,L}(u) \equiv -\mathcal J_L(u).
\end{eqnarray}
This definition will be justified by the covariant transformation law under gravitational electric-magnetic duality \eqref{transfcharges}. The entire set of celestial charges is then compactly and uniformly written as $\mathcal Q^\pm_{n,L}(u)$, $n \geq 0$. 

The multipolar decomposition of the local flux-balance laws \eqref{fb} reveals two qualitatively distinct sets: the ``memory-less'' flux-balance laws where the left-hand side of Eq.~\eqref{fb} vanishes, and the ``memory-full'' flux-balance laws where it does not.
The memory-less flux-balance laws are as follows: for $n=0$, the $\ell=0$ and $\ell=1$ scalar spherical harmonics give the standard energy and momentum flux-balance laws 
\begin{equation}
\dot{\mathcal E} \equiv \partial_u \oint_S m = \oint_S \mathcal F, \qquad \dot{\mathcal P}_i \equiv \partial_u \oint_S m n_i = \oint_S \mathcal F n_i.
\end{equation}
For $n=1$, the $\ell=1$ vector harmonics give the standard angular momentum and center-of-mass flux-balance laws \cite{Compere:2019gft}
\begin{equation}
\dot{\mathcal J}_i \equiv -\frac{1}{2}\partial_u \oint_S \mathcal N_a \epsilon^{ab}\partial_b n_i = -\frac{1}{2}\oint \mathcal F_a  \epsilon^{ab}\partial_b n_i, \qquad \dot{\mathcal K}_i \equiv  \frac{1}{2} \partial_u \oint_S \mathcal N^a \partial_a n_i = \frac{1}{2} \oint \mathcal F^a \partial_a n_i .
\end{equation}
For $n=2$ there is no memory-less flux-balance law. For $n \geq 3$, all harmonics $2 \leq \ell \leq  n-1$ give rise to memory-less flux-balance laws. For each of the two parities, there are $\sum_{\ell =2}^{n-1}(2\ell+1)=n^2-4$ such flux-balance laws. 
The $n \geq 3$ memory-less flux-balance laws are defined for  $2 \leq \ell  \leq n-1$ and take the form 
\begin{eqnarray}\label{fbNP}
\partial_u \mathcal Q^+_{n,L}(u)   = \oint_S \stackrel[(n)]{}{\mathcal F^{ab}} D_{ a} D_{b} \hat  n_L, \qquad 
\partial_u \mathcal Q^-_{n,L}(u) = \oint_S \stackrel[(n)]{}{\mathcal F^{ab}} \epsilon_{ac} D_{b} D^c \hat  n_L.
\end{eqnarray}
We will call the set of charges $\mathcal Q^+_{n,L}(u)$, $2\leq \ell \leq n-1$, $n \geq 3$ the $n \geq 3$ Newman-Penrose charges. In particular, for $n=3$ (and $\ell = 2$), one gets the 10 conserved Newman-Penrose charges \cite{1968RSPSA.305..175N}.
\footnote{Our equation \eqref{higherspincharges} for $n=3$ corresponds to Eqs. (5.48)-(5.49) of \cite{Godazgar:2018dvh} except that we use  $\stackrel[(3)]{}{\mathcal E^{ab}}$ instead of their $\stackrel[(3)]{}{ E^{ab}}$ in order to have a conserved charge in non-radiative regions.}

The $n \geq 2$ memory-full flux-balance laws defined for  $n \leq \ell$ take the form 
\begin{equation} 
\!\!\partial_u \mathcal Q^\pm_{n,L}(u)   =  \oint_S \stackrel[(n)]{}{\mathcal F^{ab}} D_{ a} D_{b} \hat  n_L + \frac{(-u)^n}{6~ n!} \oint_S \hat n_L D_\pm^{ab} \mathcal D_{n-3} \cdots \mathcal D_{0}  D_a D_c D_{\langle b}D_{d \rangle} N^{cd} , \label{fbnm}
\end{equation}
where we defined the even and odd parity second order differential operators $D_\pm^{ab}$ as $D_+^{ab}\equiv D^{\langle b} D^{a \rangle}$ and $D_-^{ab} \equiv -\epsilon^{c \langle a} D_c D^{b\rangle }$. 
The first term on the right-hand side is the ``hard'' term and the second term depending upon the moments of the news tensor is the ``soft'' or ``memory'' term. 
In non-radiative regions, the news identically vanishes. Since the right hand side of Eqs. \eqref{fbNP},\eqref{fbnm} are at least linear in the news, they also vanish by construction. The charges  $\mathcal Q^\pm_{n,L}(u)$ for any $2 \leq \ell$, $n \geq 2$ are therefore conserved in non-radiative regions. 

The linear terms in the right-hand sides of Eqs.~\eqref{fbnm} can be written as
\begin{align}
\partial_u \mathcal Q^\pm_{n,L}(u)\vert_{\text{lin}}  &=  \frac{(-u)^n}{6~ n!} \oint_S \hat n_L D_\pm^{ab} \mathcal D_{n-3} \cdots \mathcal D_{0}  D_a D_c D_{\langle b}D_{d \rangle} N^{cd}\nn \\
&= \frac{(-u)^n}{6 ~n!} \oint_S \hat n_L \prod_{k=0}^{n-3}\left(\mathcal{D}_{k}+4\alpha_k\right)D_\pm^{ab} D_a D_c D_{\langle b}D_{d \rangle} N^{cd} \nn\\
&=\frac{(-u)^n}{6 ~n!} \oint_S \hat n_L \prod_{k=2}^{n-1}\alpha_{k-2}\left[\Delta+k(k+1)\right] D_\pm^{ab} D_a D_c D_{\langle b}D_{d \rangle} N^{cd} ,
\end{align}
where we used $\mathcal{D}_n = \alpha_n (\Delta + \beta_n)$ defined in Eq.~\eqref{D_n def}. In the second equality, we have used the following commutator relations for STF tensors
\begin{align}
    [D_b,D^2]T_{a_1\cdots a_n} &= -2\sum_{s=1}^{n}\left(D_{a_s}T_{a_1\cdots a_{s-1}b a_{s+1} \cdots a_n} - \gamma_{a_s b}D^{c}T_{a_1\cdots a_{s-1}c a_{s+1} \cdots a_n}\right) -D_b T_{a_1\cdots a_n}\, ,\nn\\
    [D^{a_1},D^2]T_{a_1\cdots a_n} &=(2n-1)D^{a_1}T_{a_1\cdots a_n} . \label{CommD}
\end{align}
This algebra is facilitated by using the formula for STF tensors derived in Eq.~(39) of \cite{Toth:2021cpx} and by using a complex off-diagonal basis for the metric over the sphere. We now use the properties
\begin{subequations}
\begin{align}
D_c D_{\langle b}D_{d \rangle } N^{cd} &= D_{\langle b}D_{d \rangle }D_c N^{cd} = \frac{1}{2}D_b D_d D_c N^{cd} + \frac{1}{2} \widetilde D_b D_d D_c \widetilde N^{cd} , \label{rel8}\\ 
D^{\langle b} D^{a\rangle} D_a D_c D_{\langle b}D_{d \rangle} N^{cd}  &= \frac{1}{4}\Delta(\Delta+2) D_a D_b N^{ab}, \\
-\epsilon^{c \langle a} D_c D^{b\rangle } D_a D_c D_{\langle b}D_{d \rangle} N^{cd} &=  \frac{1}{4}\Delta(\Delta+2) D_a D_b \widetilde N^{ab},
\end{align}
\end{subequations}
where $\widetilde D^a \equiv \eps^{ab}D_b$ implying that
\begin{subequations}\label{fbnlinear2}
\begin{align}
\partial_u \mathcal Q^+_{n,L}(u)\big\vert_{\text{lin}}  &= \frac{u^n}{2^{n}~ n!} \frac{n-1}{(n+1)!}\prod_{k=0}^{n-1}\left[-\ell (\ell +1)+k(k+1)\right]  \oint_S \hat n_L D_a D_b N^{ab},
\\ 
\partial_u \mathcal Q^-_{n,L}(u)\big\vert_{\text{lin}}  &= \frac{u^n}{2^{n}~ n!} \frac{n-1}{(n+1)!}\prod_{k=0}^{n-1}\left[-\ell (\ell +1)+k(k+1)\right]  \oint_S \hat n_L D_a D_b \widetilde N^{ab}.
\end{align}
\end{subequations}
Using the convention of Eq.~\eqref{decompT}, we can further simplify $D_a D_b N^{ab}=-\Delta (\Delta+2)N^+$, $D_a D_b \widetilde N^{ab}=\Delta (\Delta+2)N^-$.

\subsection{Celestial charges in the \texorpdfstring{$Lw_{1+\infty}$}{} basis}

 The ``spin-weight'' $s$ of a tensor $(q_s)^{a_1 \dots a_k}_{b_1 \dots b_l}$ on the sphere is defined from its transformation laws under $SO(2)$, and is equal to $l-k$ \emph{i.e.}, minus the number of its contravariant indices plus the number of its covariant indices.
Following \cite{Freidel:2021ytz}, it is useful to define the spin-weight $s=-2,-1,0,1,2$ tensors 
\begin{subequations}\label{dict}
\begin{align}
 q_{-2}^{ab} &\equiv \frac{1}{2}\partial_u N^{ab}, \\ 
q_{-1}^a &\equiv \frac{1}{2}D_b N^{ab},  \\
q &\equiv \frac{1}{2}\gamma^{ab}m_{ab} +  \frac{1}{8} C_{ab} N^{ab} = m+\frac{1}{8} C_{ab}N^{ab}, \\
\widetilde{q} &\equiv \frac{1}{2}\epsilon^{ab}m_{ab} +  \frac{1}{8} C_{ab}\widetilde N^{ab} = \frac{1}{4}D_a D_b \widetilde C^{ab}+\frac{1}{8} C_{ab}\widetilde N^{ab} , \\
q_a &\equiv  N_a, \\
q_{ab} &\equiv  3 \left(\stackrel[(2)]{}{\mathcal E_{ab}} - \frac{1}{16}C_{ab}C_{cd}C^{cd}\right).
\end{align}
\end{subequations}
Such tensors are Weyl covariant  \cite{Freidel:2021qpz}\footnote{$N_a$ is denoted as $\mathcal P_a$ in \cite{Freidel:2021qpz}.}. 
We note the following two algebraic identities 
\begin{align}
\partial_u q_{-1}^a &= D_b q_{-2}^{ab} ,\label{Qam1} \\ 
\partial_u \widetilde q & = \frac{1}{2}D_a \widetilde q_{-1}^{\; a} + \frac{1}{4}C_{ab}\widetilde q_{-2}^{\; ab}.\label{Qt}
\end{align}
The first three local flux-balance laws \eqref{FBL0}-\eqref{FBL1}-\eqref{FBL2} implied by Einstein's equations can be compactly rewritten as 
\begin{align}
\partial_u q &=\frac{1}{2}D_a  q_{-1}^a + \frac{1}{4}C_{ab}q_{-2}^{ab},\label{Q} \\
\partial_u q_a &= \partial_a q+ \widetilde \partial_a  \widetilde q +C_{ab}q_{-1}^b ,\label{Qa1} \\
\partial_u q_{ab} &=D_{\langle a} q_{b \rangle}+\frac{3}{2}(C_{ab}q+\widetilde C_{ab}\widetilde q\,).\label{Qa2}
\end{align}
The tensor $ q_{0a}{}^b= q \delta_a{}^b+  \widetilde q \epsilon_{a}{}^{b}$ is spin-weight 0 since both $\delta_a{}^b$ and $\epsilon_{a}{}^{b}$ are $SO(2)$ invariants. It corresponds to the complex spin-weight 0 charge $q_0 = q+ i \widetilde q$. We note from the previous equations that $\partial_b q_{0a}{}^b = \partial_a q+\widetilde \partial_a \widetilde q$.  Equations \eqref{Qt} and \eqref{Q} can  alternatively be written as 
\begin{align}
\partial_u q_{0a}{}^b =\frac{1}{2}\delta_a^b D_c q_{-1}^c+\frac{1}{2}\epsilon_a^{\;\;b}D_c \widetilde q_{-1}^{\;c}  + \frac{1}{2} C_{ac}q^{cb}_{-2}. \label{Qa0}
\end{align}

\paragraph{Holomorphic basis.} We now define an holomorphic frame consisting of a pair of null vectors $(m^a, \overline m^a)$ on the sphere such that $m_a \overline m^a = 1$. The metric and Levi-Civita tensor can be written as $\gamma_{ab}=m_a\overline{m}_b+\overline{m}_a m_b$ and $\eps_{ab}=-i(m_a\overline{m}_b-\overline{m}_am_b)$.   
We assign $m^a$ helicity $+1$ and $\overline m_a$ helicity $-1$, so that $ T_{b_1\cdots b_l}^{a_1 \cdots a_k}\overline m_{a_l} \dots \overline m_{a_k}m^{b_1}\dots m^{b_l}$ has helicity $s=l-k$. In particular, an STF tensor $T_{a_1\cdots a_s}$, $s \geq 1$, has only two nontrivial components in this basis, with opposite helicity, given by 
\begin{align}\label{twopol}
    T_s=m^{a_1}\cdots m^{a_s}T_{a_1\cdots a_s}\,,\qquad T_{-s}=\overline m^{a_1}\cdots \overline m^{a_s}T_{a_1\cdots a_s}.
\end{align}
The shear $C=C_{ab}m^a m^b$ has helicity $+2$. Note that $\widetilde m_a = -i m_a$, $\widetilde{\overline m}_a = +i \overline m_a$.  The Geroch-Held-Penrose differential $\eth$ \cite{GHP} (which reduces in spherical coordinates to the Newman-Penrose differential \cite{1968RSPSA.305..175N}) acts on spin-weighted quantities as 
\begin{align}
\eth T &= m^{a_1} \dots m^{a_p}\overline{m}_{b_1}\dots \overline{m}_{b_q} m^b D_{b} T_{a_1\dots a_p}^{b_{1} \dots b_{q}}\,,\\
\bar{\eth} T &= m^{a_1} \dots m^{a_p}\overline{m}_{b_1}\dots \overline{m}_{b_q}\overline{m}^b D_b T_{a_1\dots a_p}^{b_{1}\dots  b_{q}}.
\end{align}
The Laplacian reads $\Delta = 2  \bar{\eth} \eth$, and the commutator is $[\bar{\eth}, \eth]T_s = s T_s $ for any spin-weighted scalar $T_s$ with helicity $s$. Note that $\eth$ and $\bar\eth$ change the helicity by $+1$ and $-1$ respectively, while $\partial_u$ does not change the helicity.

 Remarkably, the $s=-1,0,1,2$ conservation laws \eqref{Qam1}-\eqref{Qa0}-\eqref{Qa1}-\eqref{Qa2} can be expressed in the unified form \cite{Freidel:2021ytz},  
\begin{align}\label{relQ}
    \partial_u q_s = \eth q_{s-1} + \frac{s+1}{2}C q_{s-2}.
\end{align}
For the case $s=0$, Eq. \eqref{Qa0} is recovered after using 
\begin{equation}
    m^a  \overline m_b = \frac{1}{2}\delta^a_b +\frac{i}{2}\epsilon^a_{\;\;b}. 
\end{equation}

\paragraph{Higher order flux-balance laws and $Lw_{1+\infty}$ algebra.} The conjecture of \cite{Freidel:2021ytz} is that the local flux-balance laws of Einstein's equations take the form \eqref{relQ} for any $s \geq -1$ (this was proven for $s=3$ in \cite{Freidel:2021ytz}).  In particular, the flux-balance laws \eqref{relQ} for $s>2$ can be explicitly written in terms of spin-weight $s$ tensors as 
\begin{align}\label{rec}
    \partial_u q_{a_1 \cdots a_s} = D_{\langle a_1}q_{a_2 \cdots a_s\rangle } + \frac{s+1}{2}C_{\langle a_1 a_2} q_{a_3 \dots a_s \rangle}. 
\end{align}
The renormalized helicity $s$ complex charges $Q_s(\tau)$ are defined for any $s \geq 0$ as \cite{Freidel:2021dfs,Freidel:2021ytz}
\begin{align}
Q_s(\tau) \equiv \oint_S \tau_{-s}(x^a) \hat q_s(u,x^a), \qquad\hat q_s(u,x^a) \equiv \sum_{n=0}^s \frac{(-u)^{s-n}}{(s-n)!} \eth^{s-n}q_n\,. 
\end{align}
Here, the smearing function $\tau$ is a complex function over the celestial sphere.

The main result of \cite{Freidel:2021ytz} is the remarkable realization of the loop algebra $Lw_{1+\infty}$ at linear order in the radiative data. More precisely, the flux-balance laws \eqref{fb} can be used to write the $n \geq 0$ Bondi aspects as multiple retarded integrals of polynomials of (angular derivatives of) the Bondi shear and news only. One then expands these polynomials in terms of the total power of the Bondi shear and news (both counted as order 1 quantities), $Q_s(\tau) = \sum_{n \geq 1} Q^{[n]}_s(\tau)$. The linearized bracket $\{ Q_s(\tau), Q_{s'}(\tau')\}^{\text{lin}}$ is defined as $\{ Q^{[1]}_s(\tau), Q^{[2]}_{s'}(\tau')\}+\{ Q^{[2]}_s(\tau), Q^{[1]}_{s'}(\tau')\}$ since the Poisson bracket $\{\ ,\}$ \cite{Ashtekar:1981bq,Ashtekar:1981sf} decreases the power of the polynomial by 2.  At linear order, one can then obtain the bracket \cite{Freidel:2021ytz} 
\begin{align} \label{charge_algebra}
    \{ Q_s(\tau), Q_{s'}(\tau')\}^{\text{lin}}= (s'+1)Q^{[1]}_{s+s'-1}(\tau' \eth\tau)-(s+1)Q^{[1]}_{s+s'-1}(\tau \eth\tau')\,.
\end{align}

\paragraph{Dictionary between celestial charges and renormalized complex charges.}

Let us now establish the relationship between the charges $Q_s(\tau^{-s})$ for any $s \geq -2$ and the real-valued celestial charges at the linear level. From the dictionary \eqref{dict}, the relationship is clear for $s=-2,-1,0$. In order to discuss the case $s \geq 1$, we will first introduce a real basis of charges. Given a symmetric trace-free tensor $\tau^{a_1 \cdots a_s}(x^a)$, we define the real renormalized charge for any $s \geq 1$ as 
\begin{align} \label{Qs}
Q^R_{s}(\tau) = \oint_S \tau^{a_1 \cdots a_s} (x^a)\,  \Bigg(q_{a_1 \cdots a_s} &+ \sum_{n=1}^{s-1} \frac{(-u)^{s-n}}{(s-n)!} D_{a_{n+1}}  \cdots D_{a_s} q_{a_{1}\cdots a_n}  \nonumber\\
& + \frac{(-u)^{s}}{s!}D_{a_1}\cdots D_{a_{s-1}}D_{ b} q_{0 a_s }{}^b\Bigg). 
\end{align}
By convention, summed terms are removed in the absence of summation, \emph{e.g.} in the case $s=1$ here. The last term can be included as the $n=0$ term in the summation but we expand it
 given its particular index structure. We note the property $D_{\langle b}q_{0 a \rangle}^{\;\; b}= D_{ b}q_{0 a }^{\;\; b}$. For clarity, the lowest orders are given explicitly by 
\begin{align}
Q^R_1(\tau) &= \oint_S \tau^a \left(  q_a-u D_b q_{0a}{}^b \right), \\
Q^R_2(\tau) &= \oint_S \tau^{ab} \left(  q_{ab}-u D_a q_b + \frac{u^2}{2}D_a D_c q_{0b}{}^c \right), \\
Q^R_3(\tau) &=  \oint_S \tau^{abc} \left(  q_{abc}-u D_a q_{bc} + \frac{u^2}{2}D_a D_b q_c - \frac{u^3}{6} D_a D_b D_d q_{0c}{}^d \right). 
\end{align}
Using Eqs.~\eqref{rec} and \eqref{Qs} and Eq.~\eqref{rel8}, the time derivative of $Q^R_s(\tau)$ is 
\begin{equation} \label{evoleqQ}
\partial_u Q^R_s (\tau)=\frac{(-u)^s}{2 \; s!}\oint_S \tau^{a_1 \cdots a_s}D_{a_1} \dots D_{a_{s-1}} D_{\langle b} D_{ a_s \rangle }D_c N^{bc}  +      \text{non-linear terms} . 
\end{equation}
Therefore $Q^R_s(\tau)$ is conserved at linear level in non-radiative regions where the linear news identically vanishes. The relationship between the complex charge $Q_s(\tau)$  of \cite{Freidel:2021ytz} and our real basis is explicitly
\begin{align}\label{dualQ}
    Q_s(\tau)=\frac{1}{2}\left(Q_s^R(\tau)+iQ_s^R(\tl\tau)\right), 
\end{align}
from which we deduce that the charges are ``self-dual'' for any $s \geq 1$:
\begin{align}
    Q_s(\tl\tau)=-iQ_s(\tau)\,.
\end{align}
The number of polarizations of a spin-weight $s$ STF tensor $Q_s$ is 2 for any $s \geq 1$. In order to project into the two polarizations, we first define the following tensors for any $s \geq 1$, 
\begin{align}
\tau_{+,L}^{a_1 \dots a_s} \equiv D^{\langle a_1} \dots D^{a_s \rangle}\hat n_L, \qquad 
\tau_{-,L}^{a_1 \dots a_s} \equiv \widetilde D^{\langle a_1} D^{a_2} \dots D^{a_{s} \rangle } \hat n_L,
\end{align}
where $\widetilde D^a =\eps^{ab}D_b$. We further extend the definition to $s=0$ as follows: 
\begin{equation}\label{lowpolariz}
\tau_{+,L} \equiv \hat n_L,\qquad \tau_{-,L}\equiv -i \hat n_L  .
\end{equation}
We define the real-valued charges 
\begin{align}\label{QRplus}
{\mathcal Q}^{R+}_{s,L} = Q^R_{s}(\tau=\tau_{+,L}^{a_1 \dots a_s}),\qquad  {\mathcal Q}^{R-}_{s,L} = Q^R_{s}(\tau=\tau_{-,L}^{a_1 \dots a_s}).
\end{align}
We also define the helicity $-s$ quantities $\tau_{-s,\pm,L}$ as 
\begin{equation}
\tau_{-s,\pm,L}=\tau_{\pm,L}^{a_1 \dots a_s} \,\overline m_{a_1}\cdots \overline m_{a_s} ,   
\end{equation}
for $s \geq 1$. Note that $\widetilde \tau_{-s,+,L}=\tau_{-s,-,L}$, $\widetilde \tau_{-s,-,L}=-\tau_{-s,+,L}$. As a consequence of Eq.~\eqref{dualQ} we have 
\begin{eqnarray}
 Q_s(\tau_{-s,+,L}+i \tau_{-s,-,L}) = {\mathcal Q}^{R+}_{s,L}+ i {\mathcal Q}^{R-}_{s,L} , \qquad 
 Q_s(\tau_{-s,+,L}-i \tau_{-s,-,L}) = 0.
\end{eqnarray}
Equivalently, since $\mathcal Q_s$ is linear in its argument,
\begin{eqnarray}
 Q_s(\tau_{-s,+,L}) =\frac{1}{2} {\mathcal Q}^{R+}_{s,L}+ \frac{i}{2} {\mathcal Q}^{R-}_{s,L} ,\qquad 
 Q_s(\tau_{-s,-,L}) =\frac{1}{2} {\mathcal Q}^{R-}_{s,L}- \frac{i}{2} {\mathcal Q}^{R+}_{s,L} ,\label{QsR}
\end{eqnarray}
which is the multipolar analogue of Eq.~\eqref{dualQ}. Note that the complex conjugated charge $ Q^*_s(\tau_{-s,+,L}) $ is not proportional to either $ Q_s(\tau_{-s,+,L}) $ nor $Q_s(\tau_{-s,-,L}) $. 

From the dictionary \eqref{dict}, the relationship between the charges for $n=0,1,2$ defined in Eq. \eqref{QRplus} and in Eqs. \eqref{BMS01},\eqref{BMS012},\eqref{higherspincharges} at the linear level is easily found to be 
\begin{subequations}\label{loworders}
\begin{align}
{\mathcal Q}^{R+}_{0,L}&=2 \mathcal P_L = 2 {\mathcal Q}^{+}_{0,L}, &{\mathcal Q}^{R-}_{0,L}&=2 \mathcal P^-_L = 2 {\mathcal Q}^{-}_{0,L},\\ 
{\mathcal Q}^{R+}_{1,L}&=2 \mathcal K_L = 2 {\mathcal Q}^{+}_{1,L}, &{\mathcal Q}^{R-}_{1,L}&=-2 \mathcal J_L= 2 {\mathcal Q}^{-}_{1,L}, \\
{\mathcal Q}^{R+}_{2,L}&=3 \mathcal Q^+_{2,L} ,  &{\mathcal Q}^{R-}_{2,L}&=3 \mathcal Q^-_{2,L}, 
\end{align}
\end{subequations}
where we used \eqref{defQlow} in the second equality. 
Using Eqs. \eqref{BMS01}, \eqref{QsR} and \eqref{loworders}, the super-Lorentz charge $\mathcal Q_R$ associated with arbitrary infinitesimal $\text{Diff}(S^2)$ vector $Y^a=y_L^+ \partial^a \hat n_L + y_L^- \epsilon^{ab}\partial_b \hat n_L$ and the supermomenta $\mathcal Q_T$ associated with an arbitrary infinitesimal $\text{Vect}(S^2)$ scalar $T = t_L \hat n_L$ are given by 
\begin{subequations}
\begin{align}
\mathcal Q_Y &\equiv \frac{1}{2}  \oint_S Y^a \mathcal N_a =  y_L^+ \mathcal K_L - y_L^- \mathcal J_L =  \text{Re}\,  Q_1(Y^a)+\text{non-linear terms}, \\
\mathcal Q_T & \equiv \oint_S T  m = t_L \mathcal P_L= \text{Re}\,  Q_0(T)+\text{non-linear terms}.
\end{align}
\end{subequations}
We note in passing that these charges obey the generalized BMS algebra $\text{diff}(S^2) \ltimes \text{vect}(S^2)$ \cite{Barnich:2011mi,Campiglia:2015yka,Campiglia:2020qvc,Compere:2020lrt},
\begin{subequations}\label{gBMS}
\begin{align}
\{\mathcal Q_{T} , \mathcal Q_{T'}\}^{\text{lin}} &= 0,\\
\{\mathcal Q_{Y} , \mathcal Q_{T}\}^{\text{lin}} &= -\mathcal Q_{T' \equiv Y^a\partial_a T - \frac{1}{2} D_a Y^a T},\\
\{\mathcal Q_{Y} , \mathcal Q_{Y'}\}^{\text{lin}} &= -\mathcal Q_{Y''\equiv [Y,Y']}, \label{gBMS3}
\end{align}
\end{subequations}
where $[Y,Y']^a$ is the Lie bracket between $Y^a$ and $Y^{\prime a}$.

Using the formula for STF tensors derived in Eq.~(39) of \cite{Toth:2021cpx}, we can express the integrand of Eq.~\eqref{rec} 
as
\begin{align}
  \!\!  D^{\langle b} D^{a_{s-1}}\dots D^{a_1 \rangle} D_{a_1} \dots D_{a_{s-1}}  D_{\langle c} D_{b \rangle}D_{d}N^{cd} &= \frac{1}{2^{s}}\prod_{k=0}^{s-1}\left[\Delta + k(k+1)\right]D_aD_b N^{ab},\\
 \!\!    D^{\langle b}  D^{a_{s-1}}\dots D^{a_2}\widetilde D^{a_1 \rangle} D_{a_1} \dots D_{a_{s-1}} D_{\langle c} D_{b \rangle}D_{d}N^{cd} &= \frac{1}{2^{s}}\prod_{k=0}^{s-1}\left[\Delta + k(k+1)\right]D_aD_b \widetilde N^{ab}.
\end{align}
Therefore,
\begin{subequations}\label{temp009}
\begin{align}
  \partial_u {\mathcal Q}^{R+}_{n,L} &= \frac{u^n}{2^{n+1} n!} \oint_S \hat n_L \prod_{k=0}^{n-1}\left[-\ell(\ell+1) + k(k+1)\right]D_aD_b N^{ab} +      \text{non-linear terms} , \\
    \partial_u {\mathcal Q}^{R-}_{n,L} &= \frac{u^n}{2^{n+1} n!} \oint_S \hat n_L \prod_{k=0}^{n-1}\left[-\ell(\ell+1) + k(k+1)\right]D_aD_b \widetilde N^{ab} +      \text{non-linear terms} . 
\end{align}
\end{subequations}

Comparing Eqs.~\eqref{temp009} and \eqref{fbnlinear2}, we deduce for any $ n \geq 2$
\begin{equation} \label{dictionary_charge}
 \boxed{ {\mathcal Q}^{R\pm}_{n,L} = \frac{(n+1)!}{2(n-1)}  \mathcal Q^{\pm}_{n,L} + \text{non-linear terms}} 
\end{equation}
where $n=2$ reproduces \eqref{loworders}. This provides the explicit relationship between the Bondi aspects defined in \cite{Grant:2021hga} and the complex helicity $n \geq 2$ charges defined in \cite{Freidel:2021ytz} at the linear level.

While the prescriptions in \cite{Grant:2021hga} and \cite{Freidel:2021ytz} lead to conserved charges when the Bondi news vanishes, the non-linear extensions of the respective charges do not match because they are constructed from different criteria: the charges in \cite{Freidel:2021ytz} are required to be Weyl-covariant, while the charges in \cite{Grant:2021hga} are constructed with an ansatz based on the splitting of the gravitational flux into its ``hard'' and ``soft'' contributions. We shall not settle the uniqueness of the non-linear extension of the charges here. However,  we would like to emphasize that the definition of the mass aspect in \cite{Grant:2021hga} leads to the standard definition of Bondi mass $\oint_S m$ that obeys positivity theorems \cite{Schoen:1979zz,Ludvigsen:1981gf,Horowitz:1981uw} while the definition of \cite{Freidel:2021ytz} leads to $\oint_S q$ which has no definite sign.

\subsection{\texorpdfstring{$n \geq 2$}{} celestial charges from canonical multipole moments}

As just described, the local flux-balance laws allow to define two infinite sets of charges $\mathcal Q^\pm_{n,L}$, $n \geq 2$,  in non-radiative regions. We will now derive the value of these charges in the linear theory, which will allow us to give them an interpretation in terms of multipole moments. 

In the linear theory, Bondi gauge is equivalent to Newman-Unti gauge. The linear Bondi fields were obtained in \cite{Blanchet:2020ngx}.  
Using such fields in the definitions of Section \ref{sec:fbl} and defining  ${\mathcal N}_{(n),a}\equiv N_a-\frac{u}{n}D^c m_{ac}$, we obtain successively,
\begin{eqnarray}
\stackrel[(n)]{}{E_{ab}} 
&\!\!\!=\!\!\!& 4 e_{\langle a}^i e_{b \rangle}^j \frac{n-1}{n+1}\sum_{\ell \geq n} \frac{1}{\ell !} \frac{(\ell+n)!}{2^n n! (\ell-n)!}n_{L-2}\left[ M_{ijL-2}^{(\ell-n)} +\frac{2\ell}{\ell+1}\epsilon_{ipq} n_p S_{jqL-2}^{(\ell-n)}\right]+\mathcal{O}(G), \nonumber\\
{\mathcal N}_{(n),a}
&\!\!\!=\!\!\!& e^i_a \sum_{\ell \geq 1} \frac{(\ell+1)(\ell+2)}{2(\ell-1)!}n_{L-1} \left(1-\frac{u}{n}\partial_u \right) \left[ M_{iL-1}^{(\ell-1)} + \frac{2\ell}{\ell+1}\epsilon_{i p q}n_p S_{qL-1}^{(\ell-1)}\right]+\mathcal{O}(G),\\
D_{\langle a} {\mathcal N}_{(n),b\rangle } &\!\!\!=\!\!\!& e_{\langle a}^i e_{b \rangle}^j \sum_{\ell \geq 2}  \frac{(\ell+1)(\ell+2)}{2(\ell-2)!}n_{L-2} \left(1-\frac{u}{n}\partial_u \right)\left[ M_{ijL-2}^{(\ell-1)} + \frac{2\ell}{\ell+1}\epsilon_{i p q}n_p S_{qjL-2}^{(\ell-1)}\right]+\mathcal{O}(G).\nonumber
\end{eqnarray}
Upon acting with $\Delta$ on $D_{\langle a} {\mathcal N}_{(n),b\rangle }$ we can use 
\begin{eqnarray}
(\Delta +\ell^2+\ell-4)\left[e_{\langle a}^i e_{b\rangle}^j n_{L-2}(T^+_{ijL-2}+\epsilon_{ipq}n_pT^-_{jqL-2})\right]=0. 
\end{eqnarray}
We also note the properties
\begin{subequations}
\begin{align}
D^b \left[ e_{\langle a}^i e_{b \rangle }^j n_{L-2} (X^+_{ijL-2}+\epsilon_{ipq} n_p X^-_{jqL-2}) \right] &= -\frac{\ell+2}{2} e^i_a n_{L-1} ( X^+_{iL-1}+\epsilon_{ipq}n_p X^-_{q L-1})   ,  \\ 
D^a \left[e^i_a n_{L-1} (X^+_{iL-1}+\epsilon_{ipq}n_p X^-_{q L-1})\right] &= -(\ell+1)n_L X^+_L,\label{props} \\
\epsilon^{ba}D_b \left[e^i_a n_{L-1} (X^+_{iL-1}+\epsilon_{ipq}n_p X^-_{q L-1})\right] &= -(\ell+1) n_L X^-_L,
\end{align}
\end{subequations}
for any STF tensors $X^+_L$, $X^-_L$. 
Substituting this expression in Eq.~\eqref{higherspincharges}, using 
\begin{align}
T_L {T^\prime}_{L'} \oint_S n_L n_{L'} =T_L {T^\prime}_L\delta_{\ell,\ell'}\ell!/(2\ell+1)!!    
\end{align}
for STF tensors $T_L,\, {T^\prime}_{L'}$, we deduce for each $n \geq 2$ 
\begin{align}
\mathcal Q^+_{n,L}(u)  &= \left\{ \begin{array}{ll} a_{n,\ell} M_{L}^{(\ell-n)}+ \sum_{p=0}^{n-3}q_{n,\ell,p} u^{p+1}M^{(\ell-n+p+1)}_L  & \\ \qquad +b_{n,\ell}u^{n-1}\left(1-\frac{u}{n}\partial_u\right)M^{(\ell-1)}_L  +\mathcal{O}(G),  & \qquad \ell \geq n, \\ \newline{}\\ \sum_{p=n-\ell-1}^{n-3}q_{n,\ell,p} u^{p+1}M^{(\ell-n+p+1)}_L & \\\qquad  +b_{n,\ell}u^{n-1}\left(1-\frac{u}{n}\partial_u\right)M^{(\ell-1)}_L + \mathcal{O}(G), & \qquad 2\leq \ell  \leq n-1,\end{array}\right. \label{QpnL}
\end{align}
where 
\begin{subequations}\label{coefficients}
\begin{align}
d_{n,\ell} &= -\frac{n+2}{2(n+1)(n+4)} \Big((n+2)(n+3)-\ell(\ell+1)\Big), \label{dn}\\
a_{n,\ell} &= \frac{(\ell+2)(\ell+1) (n-1)(\ell+n)!}{2^{n-1}(n+1)! (\ell-n)! (2\ell+1)!!} ,\label{an} \\ 
b_{n,\ell} &=(-1)^{n-1} \frac{(\Pi_{m=0}^{n-3} d_{m,\ell}) (\ell+2)^2(\ell+1)^2\ell(\ell-1)}{12(n-1)!(2\ell+1)!!},\label{bn}\\
q_{n,\ell,p} &= \frac{(-1)^{p+1}}{(p+1)!} a_{n-p-1,\ell} \left(\Pi_{m=0}^{p} d_{n-m-3,\ell}\right).\label{qn}
\end{align}
\end{subequations}
The expression for $\mathcal Q^-_{n,L}(u) $ is identical to \eqref{QpnL} with $M_L$ replaced with $M^-_L = \frac{2\ell}{\ell+1}S_L$, consistently with gravitational electric-magnetic duality, see Section \ref{EMD}. Note that $q_{n,\ell,-1}=a_{n,\ell}$. In summary, we obtained the exact expression of the $n \geq 2$ celestial charges $\mathcal Q^\pm_{n,L}(u)$ in terms of canonical multipole moments in the linear theory. We can therefore equally call the celestial charges $\mathcal Q^\pm_{n,L}$ as the $n \geq 2$  celestial multipoles.

\section{Gravitational electric-magnetic duality}\label{EMD}

The linearized Einstein-Hilbert action admits a $SO(2)$ duality symmetry under the rotation 
\begin{equation}\label{duality}
\left( \begin{array}{c} R_{\alpha\beta\mu\nu}\\  {}^\star R_{\alpha\beta\mu\nu}\end{array}\right)  \mapsto   \left(\begin{array}{cc}  \cos\psi & \sin\psi \\ -\sin\psi& \cos\psi \end{array} \right) \left( \begin{array}{c} R_{\alpha\beta\mu\nu}\\  {}^\star R_{\alpha\beta\mu\nu}\end{array}\right) \equiv  R(\psi)  \left( \begin{array}{c} R_{\alpha\beta\mu\nu}\\  {}^\star R_{\alpha\beta\mu\nu}\end{array}\right),
\end{equation}
where the dual Riemann tensor is defined as ${}^\star R_{\alpha\beta\mu\nu}\equiv \frac{1}{2}\epsilon_{\alpha\beta\gamma\delta}R^{\gamma\delta}_{\;\;\;\mu\nu}$ \cite{Henneaux:2004jw}. This duality transformation is formulated in terms of tensors and is therefore independent of the choice of coordinates. Let us take as a boundary condition the absence of incoming gravitational radiation. In harmonic coordinates denoted by $\bar x^\mu=(\bar t,\bar x^i)$ with retarded time $\bar u = \bar t - \bar r$, the most general linearized metric, up to residual gauge transformations, takes the  canonical form \cite{Thorne:1980ru,Blanchet:2013haa} 
\begin{subequations}\label{eq:linearmetric}
\begin{align}
    \bar h^{00} &= 4 \sum_{\ell=0}^{+\infty}\frac{(-)^{\ell}}{\ell!}\bar\p_L\left(\frac{M_{L}(\bar u)}{\bar r} \right)\,, \\
    \bar h^{0j} &= -4\sum_{\ell=1}^{+\infty} \frac{(-)^{\ell}}{\ell!}\left[\bar\p_{L-1}\left(\frac{M^{(1)}_{j L-1}(\bar u)}{\bar r} \right) + \frac{\ell}{\ell+1}\bar\p_{p L-1}\left(\frac{\varepsilon_{jpq}S_{q L-1}(\bar u)}{\bar r} \right)\right]\,, \\
    \bar h^{jk} &=  4\sum_{\ell=2}^{+\infty} \frac{(-)^{\ell}}{\ell!}\left[\bar\p_{L-2}\left(\frac{M^{(2)}_{jk L-2}(\bar u)}{\bar r} \right) +  \frac{2\ell}{\ell+1}\bar\p_{p L-2}\left(\frac{\varepsilon_{pq(j}S^{(1)}_{k)q L-2}(\bar u)}{\bar r} \right)\right]\,,
\end{align}
\end{subequations}
where $\bar h^{\mu\nu}=h^{(1)\mu\nu}-\frac{1}{2}\eta_{\alpha\beta}h^{(1)\alpha\beta} \eta^{\mu\nu}$ is the trace-reversed metric perturbation, $h^{(1)}_{\mu\nu}$ is the metric perturbation, $\bar\p_\mu=\partial/\partial\bar x^\mu$ and superscripts $(i)$ on $M_L$, $S_L$, $i=1,2,\dots $ denote the number of $\bar u$ derivatives, \emph{i.e.} $X^{(n)}\equiv {\pd_{\bar u}}^n X$. The corresponding metric in Bondi coordinates $(u,r,\th^a)$ reads up to a BMS transformation as \cite{Blanchet:2020ngx} 
\begin{subequations}\label{eq:metric}
\begin{align}
 g_{uu} &= -1 + 2G \sum_{\ell=0}^{+\infty}\frac{(\ell+1)(\ell+2)}{\ell!}\sum_{k=0}^{\ell}\frac{\alpha_{k\ell}}{(k+1)(k+2)} \frac{n_L M_L^{(\ell-k)}}{r^{k+1}}\,,\label{eq:metricuu}\\
 g_{ua} &= G\,e_a^i\biggl\{ - \sum_{\ell=2}^{+\infty}\frac{\ell+2}{\ell!}n_{L-1}\Bigl[  M_{iL-1}^{(\ell)} - \frac{2\ell}{\ell+1}\varepsilon_{ipq}n_p S_{qL-1}^{(\ell)}\Bigr] \nn\\
& \qquad+ 2\sum_{\ell=1}^{+\infty}\frac{\ell+2}{\ell!}n_{L-1}\sum_{k=1}^{\ell}\frac{\alpha_{k\ell}}{k+2} \frac{1}{r^k} \Bigl[M_{iL-1}^{(\ell-k)}+\frac{2\ell}{\ell+1}\varepsilon_{ipq}n_p S_{qL-1}^{(\ell-k)}\Bigr]\biggr\}\,,\label{eq:metricua}\\
 g_{ab} &= r^2 \Biggl[ \gamma_{ab} + 4G\,e_{\langle a}^i e_{b\rangle}^j \sum_{\ell=2}^{+\infty}\frac{1}{\ell!}\frac{n_{L-2}}{r}\biggl\{ M_{ijL-2}^{(\ell)}-\frac{2\ell}{\ell+1}\varepsilon_{ipq}n_p S_{jqL-2}^{(\ell)} \nn\\
&\qquad\qquad\qquad 
+ \sum_{k=2}^{\ell}\frac{k-1}{k+1} \frac{\alpha_{k\ell}}{r^k} \Bigl[M_{ijL-2}^{(\ell-k)}+\frac{2\ell}{\ell+1}\varepsilon_{ipq}n_p S_{jqL-2}^{(\ell-k)}\Bigr]\biggr\} \Biggr] \,,\label{eq:metricab}
\end{align}
\end{subequations}
where $\alpha_{n k}=\frac{(k+n)!}{2^n n! (k-n)!}$. The metric is functionally dependent upon the two sets of canonical multipole moments that are conveniently defined as 
\begin{align}\label{ML}
M_L^+ (u) = M_L(u), \qquad M_L^-(u)  = \frac{2\ell}{\ell+1}S_L(u)
\end{align}
in terms of the retarded time $u$. The Riemann tensor of that linearized metric can be thought of as a functional of the canonical multipole moments $M^\pm_L(u)$ as well as of the coordinates, \emph{i.e.}, $R_{\alpha\beta\mu\nu} = R_{\alpha\beta\mu\nu}(M^\pm_L(u) ; x^\mu)$. The duality symmetry implies that the dual Riemann tensor of a linearized solution is equal to the Riemann tensor of another linearized solution, which also takes a canonical form \eqref{eq:linearmetric} or \eqref{eq:metric}. 
Using a computer algebra software, we obtain that
\begin{equation}
 {}^\star R_{\alpha\beta\mu\nu}(M^+_L(u),M^-_L (u); x^\mu)  =  R_{\alpha\beta\mu\nu}(M^-_L(u), -M^+_L(u) ; x^\mu) ,
\end{equation}
\emph{i.e.} the dual linearized Riemann tensor is the linearized Riemann tensor of the metric $g^{\text{lin}}_{\mu\nu}(M^-_L(u),-M^+_L (u); x^\mu)$ with dualized moments $M^+_L (u)\mapsto M^-_L(u)$, $M^-_L(u) \mapsto -M^+_L(u)$. Under a finite duality rotation \eqref{duality} the multipole moments are transformed as 
\begin{equation}\label{actionML}
\left( \begin{array}{c} M^+_L \\  M^-_L \end{array}\right)\mapsto \left( \begin{array}{c} \mbox{}^\star M^{+}_L \\  \mbox{}^\star M^{-}_L \end{array}\right)\equiv R(\psi) \left( \begin{array}{c} M^+_L \\  M^-_L \end{array}\right)=\left(\begin{array}{cc} \cos\psi & \sin\psi \\ -\sin\psi& \cos\psi \end{array} \right)\left( \begin{array}{c} M^+_L \\  M^-_L \end{array}\right).
\end{equation}
In the multipolar post-Minkowskian scheme \cite{Blanchet:2013haa}, the most general non-linear metric solution of vacuum Einstein's equations with no incoming radiation is uniquely constructed, up to gauge transformations, from the linear metric $g^{\text{lin}}_{\mu\nu}(M^+_L(u), M^-_L (u); x^\mu)$. Therefore, the linear duality symmetry \eqref{actionML} maps a full solution of general relativity to another one. This way, we can promote the action of the linear duality symmetry \eqref{actionML} of the linear solution space onto the non-linear solution space with no incoming radiation. More precisely, the 
non-linear perturbative solution of Einstein's equations in the exterior zone outside of sources reads as $g_{\mu\nu}(M_L^+,M_L^-; x^\mu)=\eta_{\mu\nu}+\sum_{n=1}^\infty G^n h^{(n)}_{\mu\nu}$ where $h^{(n)}_{\mu\nu}$, $n \geq 2$ are built from the post-Minkowskian algorithm from $h^{(1)}_{\mu\nu}(M_L^+,M_L^-; x^\mu)$ for any $M_L^\pm(u)$. Given a set $M_L^\pm(u)$ we define the dual metric as $g_{\mu\nu}({}^\star M_L^{+,\text{n.l.}},{}^\star M_L^{-,\text{n.l.}}; x^\mu)$ where ${}^\star M_L^{\pm,\text{n.l.}}(u)$ are the \emph{non-linear dual moments} 
\begin{equation}
{}^\star M_L^{\pm,\text{n.l.}}(u) =   {}^\star M_L^{\pm}(u) + G  \stackrel[(1)]{}{{}^\star M_L^{\pm}}(u) + \mathcal{O}(G^2)  .\label{defdual}
\end{equation}

The linear terms are defined from the right-hand side of Eq.~\eqref{actionML}. The dual metric is a perturbative solution simply because the post-Minkowskian algorithm is defined for any sets of functions $M_L^\pm(u)$ and, in particular, for the dual set ${}^\star M_L^{\pm,\text{n.l.}}(u)$. This proves that the duality \cite{Henneaux:2004jw} extends to the non-linear solution space considered for any choice of subleading corrections $\mathcal{O}(G)$ in \eqref{defdual}.\footnote{We can further extend the duality in the presence of incoming radiation after linearly superposing to $g^{\text{lin}}_{\mu\nu}(M_L^\pm(u) ; x^\mu)$ the incoming solution parametrized by a second set of canonical multipole moments $M_L^\pm(v)$ that are functions of advanced time. The non-linear perturbative metric in $G$ has not been constructed to our knowledge but by construction is functionally dependent upon the four sets of multipoles which are exchanged as pairs under duality.}
We will prove that there is a nonlinear definition of the duality such
that it is a symmetry of the symplectic structure at null infinity. However, we don't show that transformation is a symmetry of the action.

At linear level, the shear is transformed under duality as 
\begin{align}
\left( \begin{array}{c}  C_{ab}(u,\theta^a) \\  \widetilde C_{ab}(u,\theta^a) \end{array}\right)\mapsto  \left( \begin{array}{c} {}^\star C_{ab}(u,\theta^a) \\   {}^\star\widetilde C_{ab}(u,\theta^a) \end{array}\right)\equiv R(\psi) \left( \begin{array}{c}  C_{ab}(u,\theta^a) \\  \widetilde C_{ab}(u,\theta^a) \end{array}\right), \label{dualCpsi}
\end{align}
where $R(\psi)$  is a duality rotation. The transformation of the news follows by differentiation with respect to $u$. Let us now promote the duality transformation \eqref{dualCpsi} to the non-linear theory. In the terminology of \cite{Blanchet:2013haa}, Eq.~\eqref{dualCpsi} amounts to defining dual radiative multipole moments ${}^\star U_L(u)$, ${}^\star V_L(u)$ as a function of $M_L,S_L$ and the angle $\psi$. The functional definition of the radiative multipole moments  in terms of the canonical multipole moments $U^\pm_L=U^\pm_L(M_L^\pm)$ reads as (see, \emph{e.g.}, Eqs. (88) of \cite{Blanchet:2013haa} with $U_L^- \equiv \frac{2\ell}{\ell+1}V_L$),
\begin{align}
U^\pm_L &= M^{\pm(\ell)}_L + G \stackrel[(1)]{}{U^\pm_L} + \mathcal{O}(G^2).
\end{align}
We then define the $\mathcal{O}(G)$ corrections to the dual canonical multipole moments ${}^\star M_L^{\pm,\text{n.l.}}$ from solving the equations ${}^\star U^\pm_L=U^\pm_L({}^\star M_L^{\pm,\text{n.l.}})$. For $\psi$ infinitesimal, we can expand the duality operation as $\star = I+\psi \delta_\star+O(\psi^2)$ where $I$ is the identity and $\delta_\star$ is the differential operator that infinitesimally varies $M_L$ into the dual ${}^\star M_L$ and we truncate to linear order in $\psi$. 
The equation at order $G$ takes the form 
\begin{align}
    \stackrel[(1)]{}{{}^\star M_L^{\pm (\ell)}}(u) &= \pm \big(\stackrel[(1)]{}{U^\mp_L}(u)- \delta_{\star} \stackrel[(1)]{}{U^\pm_L}(u)\big) .
\end{align}
Qualitatively similar equations are obtained at higher order in $G$. These equations admit a unique (non-local) solution for the corrections $ \stackrel[(n)]{}{{}^\star M_L^{\pm}}$, $n \geq 1$ up to integration constants which are nothing else than the higher $G$ corrections to the celestial charges in the non-radiative asymptotic region, see Section \ref{sec:cl}. We complete the definition of the duality by imposing that the conserved charges at past infinity transform as
\begin{equation}\label{transfcharges}
\left( \begin{array}{c} \mathcal Q^+_{n,L} \\  \mathcal Q^-_{n,L} \end{array}\right)\mapsto   R(\psi)  \left( \begin{array}{c} \mathcal Q^+_{n,L} \\  \mathcal Q^-_{n,L} \end{array}\right),
\end{equation}
for any $n \geq 0$ after using the definition \eqref{defQlow} for the cases $n=0,1$. Eqs.~\eqref{dualCpsi} and \eqref{transfcharges} uniquely define the duality $SO(2)$ rotation in the non-linear theory. 

Since the Bondi-Sachs expansion at any order $n$ admits at most interactions of order $\text{max}(2,n+1)$, the non-linear duality symmetry induced by the $SO(2)$ rotation of canonical multipole moments \eqref{actionML} acts at each order in the Bondi-Sachs expansion. At leading order in the large radius expansion, the dual of the $n=0$ flux-balance law \eqref{FBL0} is given by 
\begin{equation}
\frac{1}{4}D_b D_c \widetilde N^{bc} =- {\mathcal F}^- +\partial_u m^- ,
\end{equation}
where the flux identically vanishes $ {\mathcal F}^- \equiv 0$ since $N_{ab}\widetilde N^{ab}=0$. This coincides with Eq.~\eqref{Qt} for the dual mass aspect. Einstein equations are therefore covariant under duality at leading order in the Bondi expansion. We expect that at each higher order in the Bondi expansion the flux-balance laws will similarly transform covariantly under duality. This remains to be explicitly derived.  If this is true, we expect that the transformation law of the charges \eqref{transfcharges} will hold not only at past null infinity by construction but at any $u$. This remains to be investigated.

The above symmetry transformation can be generalized by making the rotation angle dependent upon the individual harmonic $L$, \emph{i.e.}, $\psi = \psi (\theta^a)$. Such transformation still maps a linearized solution to another one, which again can be promoted to a transformation of the non-linear solution space as described above. This transformation induces (through \eqref{shear multipole expansion} below) a local transformation of the shear and the news on the celestial sphere
\begin{align}
&\left( \begin{array}{c}  C_{ab}(u,\theta^a) \\  \widetilde C_{ab}(u,\theta^a) \end{array}\right)\mapsto R(\psi(\theta^a)) \left( \begin{array}{c}  C_{ab}(u,\theta^a) \\  \widetilde C_{ab}(u,\theta^a) \end{array}\right),  
\left( \begin{array}{c}  N_{ab}(u,\theta^a) \\  \widetilde N_{ab}(u,\theta^a) \end{array}\right) \mapsto R(\psi(\theta^a)) \left( \begin{array}{c}  N_{ab}(u,\theta^a) \\  \widetilde N_{ab}(u,\theta^a) \end{array}\right),
\end{align}
where $R(\psi(\theta^a))$  is a duality rotation that depends on the angles on the celestial sphere.

As was shown in \cite{Seraj:2021rxd,Seraj:2022qyt}, this localized transformation is a symmetry of the symplectic structure of the radiative phase space at null infinity, \emph{i.e.}
\begin{align}
    \cL_{\varepsilon} \Omega=0,\qquad \Omega=\frac{1}{8G}\int du \oint_{S^2}\delta N^{ab} \wedge\delta C_{ab}.
\end{align}
Here $\cL_{\varepsilon}=\delta i_{\varepsilon}+i_\varepsilon \delta$ is the phase space Lie derivative along a vector field corresponding to the field variation $\de_\varepsilon$ and $\delta_\varepsilon C_{ab} = i_\varepsilon \delta C_{ab} = \varepsilon(\theta^a)\widetilde C_{ab}$ where the rotation matrix is infinitesimally expanded as $(R(\psi(\theta^a))_{ab}=\delta_{ab}+\varepsilon(\theta^a)\epsilon_{ab}$. This symmetry is therefore canonically generated by a Hamiltonian charge, which naturally appears in the gyroscopic gravitational memory \cite{Seraj:2021rxd,Seraj:2022qyt}.

While gravitational electric-magnetic duality is a symmetry of the  solution space of Einstein gravity constructed from arbitrary canonical multipoles $M_L^\pm$, this solution space contains solutions that are usually discarded by boundary conditions. In particular, Taub-NUT charges are ruled out because of the presence of closed timelike curves \cite{1963JMP.....4..924M}. For standard boundary conditions \cite{Strominger:2013jfa}, the shear is imposed to be purely electric at early and large retarded times $u \rightarrow \pm \infty$ in accordance with the soft graviton theorem \cite{He:2014laa}. These two conditions amount to the vanishing of the dual Bondi supermomenta at early and late times: 
\begin{equation}\label{symmbreaking}
  \text{lim}_{u \rightarrow \pm \infty} \mathcal P^-_L = 0\quad  \leftrightarrow \quad S_{L,\ell}=0, \quad \forall L, \; \ell \geq 0.
\end{equation}
The second equality follows after using Eq. \eqref{PD} since spacetime is non-radiative as $u \rightarrow \pm \infty$. Restricting the phase space by imposing \eqref{symmbreaking} breaks the duality symmetry. Extended boundary conditions that allow non-vanishing dual Bondi supermomenta $(\ell \geq 2)$, while keeping the NUT charges zero $(\ell=0,1)$, can also be considered \cite{Choi:2019sjs}, as they might accommodate physically relevant solutions \cite{Satishchandran:2019pyc}.

\section{Complete set of conserved quantities for non-radiative spacetimes}
\label{sec:cl}

We will now prove in the linear theory and argue in the non-linear theory that the non-radiative spacetimes are entirely characterized by the data of the celestial charges excluding the Newman-Penrose charges, which are derived quantities. We will call the complete set of non-radiative data the non-radiative multipole charges and denote them as $M_{L,k}^\pm$, $0 \leq k \leq \ell$.

In regions where the news vanishes, we have $M_L^{(\ell+1)}=S_L^{(\ell+1)}=0$ in the linear theory and therefore the linear canonical multipole moments admit the expansion 
\begin{eqnarray}\label{expansion}
M_L(u) = \sum_{k=0}^\ell M_{L,k} u^k, \qquad S_L(u) = \sum_{k=0}^\ell S_{L,k} u^k, 
\end{eqnarray}
which implies $M_L^{(\ell-n)}(u)=\sum_{k=0}^n M_{L,k+\ell-n}\frac{(k+\ell-n)!}{k!}u^k$. A non-radiative linear metric  is therefore uniquely characterized by the set of constants $M_{L,k},S_{L,k} $ with integer labels $\ell\geq 0$ and $0 \leq  k\leq \ell$. In the case of a transition between a non-radiative region at $u \leq u_i$ to another non-radiative region $u \geq u_f$, the expansion \eqref{expansion} is only valid at early and late times. In the simple case of scattering of massive particles, the canonical multipole moments $M_L$, $S_L$ can be expressed in terms of the initial positions and momenta of the incoming or outgoing particles. In general, the quantities $M_{L,k} $, $S_{L,k}$ are functionals of the incoming and outgoing states.

The coefficients $M_{L,0},\; S_{L,0}$ encode, up to a normalization coefficient, the stationary Geroch-Hansen multipole moments, while the coefficients $M_{L,k},\; S_{L,k}$ for $1 \leq k \leq \ell$ encode non-radiative non-stationary features. In particular, $M_{L,\ell},\, S_{L,\ell}$ for $\ell \geq 2$ encode the time-independent shear
\begin{align}\label{shear multipole expansion}
    C_{a b}=4e_{\langle a}^{i} e_{b\rangle}^{j} \sum_{\ell=2}^{+\infty} n_{L-2}\left[M_{ij L-2,\ell}-\frac{2 \ell}{\ell+1} \varepsilon_{i p q} n_{p} S_{j q L-2,\ell}\right].
\end{align}
No known astrophysical source generates a constant odd-parity shear and the charges $S_{L,\ell}$ for $\ell \geq 2$ are assumed to be vanishing for standard boundary conditions as already stated in Eq. \eqref{symmbreaking} (see however \cite{Satishchandran:2019pyc}). The difference of charges $M_{L,\ell}$ for $\ell \geq 2$ between an initial $u=u_i$ and final $u=u_f$ non-radiative region equivalently encodes the (linear) displacement memory. This directly follows from integrating the $n=0$ flux-balance law \eqref{FBL0} between $u=u_i$ and $u=u_f$. The charges $M_{L,k},S_{L,k} $, $1 \leq k \leq \ell-1$ encode further non-radiative non-stationary features that are independent from the displacement memory effect since they do not affect the linear shear. The difference of the value of these charges at early and late times in non-radiative regions encodes, by uniqueness, the subleading memory effects \cite{Pasterski:2015tva,Nichols:2017rqr,Nichols:2018qac,Flanagan:2018yzh,Compere:2019odm,Grant:2021hga,Seraj:2021rxd,Seraj:2022qyt}. This can be deduced from integrating Eqs. \eqref{FBL1}- \eqref{FBL2}- \eqref{FBL3} between $u=u_i$ and $u=u_f$. Since the linear news is zero in non-radiative regions, one can integrate the left-hand side of each flux-balance law by parts and obtain an equality of the type $[\mathcal Q_{n,L}^\pm ]^{u=u_f}_{u=u_i} \sim [M^\pm_{L,\ell-n}]^{u=u_f}_{u=u_i}+\mathcal{O}(G)$ for all $n \geq 1$ after using Eq. (27) of \cite{Blanchet:2020ngx}. The proportionality factor will be computed in the following. This indicates that $M^\pm_{L,\ell-n}$, $n \geq 1$ are precisely the charges encoding the subleading memory effects.

We now derive the value of the $n \geq 2$ celestial charges $\mathcal Q^\pm_{n,L}$ \eqref{higherspincharges} in terms of the coefficients $M_{L,k},\; S_{L,k}$. Using Eq. \eqref{QpnL} we obtain 
\begin{equation}
\boxed{\mathcal Q^\pm_{n,L}=0+\mathcal{O}(G), \qquad 2 \leq \ell \leq n-1}
\end{equation}
and
\begin{equation}\label{23}
\boxed{\mathcal Q^\pm_{n,L}= a_{n,\ell} (\ell-n)! M^\pm_{L,\ell-n}+\mathcal{O}(G) ,\qquad \ell \geq n,}
\end{equation}
where the duality-covariant notation \eqref{ML} makes the final result elegant. This provides a clear and physical interpretation of the conserved $n \geq 2$ celestial charges. There are two qualitatively distinct sets defined in the range $2 \leq \ell \leq n-1$ and $\ell \geq n$. For $\ell \geq n$, the conserved $n \geq 2$ celestial charges  parametrize non-radiative non-stationary features of the gravitational field at null infinity. While the stationary Geroch-Hansen multipole moments $M_{L,0}^\pm$, $\ell \geq 0$, first appear in the large radius expansion at order $r^{-\ell-1}$ in $g_{uu}$, $r^{-1} g_{ua}$ and $r^{-2} g_{ab}$, the non-stationary non-radiative multipole moments $M_{L,k}^\pm$ first appear at $r^{k-\ell-1}$ in $g_{uu}$, $r^{-1} g_{ua}$ and $r^{-2} g_{ab}$ for any $1 \leq k \leq \ell$. The cases $2 \leq \ell \leq n-1$ where all charges identically vanish in the linear solution space considered   exactly correspond to the memory-less flux-balance laws. In particular, the 10 charges $\mathcal Q^\pm_{3,ij}$ for $n=3$, $\ell=2$ are precisely the $n=3$ Newman-Penrose charges which vanish at linear level in $G$ consistently with known quadratic expressions in terms of multipole moments \cite{10.2307/2415610}.

The fact that our explicit computation gives conserved charges independent of $u$ also validates the $u$-renormalized quantity \eqref{Eabn} of Grant-Nichols \cite{Grant:2021hga} at the linear level, and it provides a non-trivial cross-check of the coefficients \eqref{coefficients}.

The conserved charges $\mathcal Q^\pm_{n,L}$, $\ell \geq n \geq 2$ provide the quantities $M_{L,k}$, $S_{L,k}$ for all $\ell \geq 2$ and $0 \leq k \leq \ell-2$, which is not the complete set of quantities in Eq. \eqref{expansion}. In fact, the complementary set is given by the generalized BMS charges \eqref{BMS01}, including the dual supermomenta \eqref{BMS012}. 
For a non-radiative region, using Eq.~\eqref{expansion} and the expressions Eq.~(25) of \cite{Blanchet:2020ngx} that we report here for the ease of the presentation,
\begin{subequations}
	\begin{align}
	m &= \sum_{\ell=0}^{+\infty} \dfrac{(\ell+1)(\ell+2)}{2\ell!} \,n_L M_{L}^{(\ell)} + \mathcal{O}(G)\,,\label{eq:maspect}\\
	N_{a} &= e_{a}^i \sum_{\ell=1}^{+\infty}\frac{(\ell+1)(\ell+2)}{2(\ell-1)!} \,n_{L-1}\left[ M_{iL-1}^{(\ell-1)}+\frac{2\ell}{\ell+1}\varepsilon_{ipq}n_p S_{qL-1}^{(\ell-1)}\right] + \mathcal{O}(G)\,,
	\end{align}
\end{subequations}
the Bondi supermomenta and super-Lorentz charges at the leading order in $G$ read as  
\begin{align}
\mathcal P_L &= \frac{ (\ell+2)!}{2(2\ell+1)!!}M_{L,\ell},\qquad &\ell &\geq 0,\label{PLexpr}\\ 
\mathcal J_L&=-\frac{\ell^2(\ell+1)(\ell+2)}{ 2(2\ell+1)!!}(1-u \partial_u) S_L^{(\ell-1)}=-\frac{\ell(\ell+2)!}{ 2(2\ell+1)!!} S_{L,\ell-1}, \qquad &\ell &\geq 1,\label{JLexpr}\\ 
\mathcal K_L &=\frac{\ell(\ell+1)^2(\ell+2)}{ 4 (2\ell+1)!!}(1-u \partial_u) M_L^{(\ell-1)}=\frac{(\ell+1)(\ell+2)!}{ 4 (2\ell+1)!!} M_{L,\ell-1},\qquad &\ell& \geq 1.\label{KLexpr}
\end{align}

Using Eq. (27) of \cite{Blanchet:2020ngx} and the relations \eqref{props}, the dual Bondi supermomenta \eqref{BMS012}  are obtained as 
\begin{align}\label{PD}
{\mathcal P}^-_L = \frac{1}{2}\oint_S m_{ab}\epsilon^{ab} \, \hat n_L = \oint_S m^-\,  \hat n_L = \frac{\ell (\ell+2)!}{(\ell+1)(2\ell+1)!!}S_{L,\ell},\qquad \ell \geq 1. 
\end{align}
As a part of standard boundary conditions \eqref{symmbreaking}, the dual supermomenta vanish in non-radiative regions. As a non-trivial check of the coefficients appearing in \eqref{PLexpr}, \eqref{JLexpr}, \eqref{KLexpr}, \eqref{PD}, one can verify that the transformation laws under gravitational electric-magnetic duality \eqref{transfcharges} are obeyed, see  Section \ref{EMD}.

Let us now argue that the set of charges $M^\pm_{L,k}$, $0 \leq k \leq \ell$ completely characterizes a non-radiative spacetime (at all retarded times $u$) including perturbative interactions. A generic non-linear metric perturbatively constructed from the post-Minkowskian expansion with no incoming gravitational wave flux is entirely determined by the linear canonical multipole moments $M^\pm_L(u)$. If the non-linear terms in the news tensor vanish when the  $\ell+1$ derivatives of $M^\pm_L(u)$ vanish, the linear canonical multipole moments given by Eq. \eqref{expansion} still parameterize a generic non-radiative solution. It is known to be the case for the tail interactions and hereditary terms in the quadrupole-quadrupole approximation \cite{Blanchet:2020ngx}. 
Assuming this behaviour remains valid for any interactions, we have now demonstrated the following completeness relation. The set of charges $\mathcal Q_{n,L}^\pm$, $\ell \geq n \geq 2$ together with $\mathcal P_L$, $\ell \geq 0$ and $\mathcal J_L$, $\mathcal K_L$, $\ell \geq 1$ are complete in the sense that they span the entire phase space of non-radiative spacetimes in the post-Minkowskian expansion. For the ease of the reader, we provide in Tables \ref{tab:1}, \ref{tab:2} the list of the sets of conserved charges for $\ell \leq 3$ which can be extended straightforwardly to arbitrary $\ell$.

\begin{table}[!htb]
	\centering
	\begin{tabular}{|c|c|c|c|}\hline
		$M_{\emptyset,0}\sim \mathcal E $& & & \\ \hline
		$ M_{i,0} \sim \mathcal K_i  $& $ M_{i,1} \sim \mathcal P_i  $& & \\   \hline 
		$ M_{ij,0} \sim \mathcal Q^+_{2,ij}  $& $ M_{ij,1} \sim \mathcal K_{ij}  $& $ M_{ij,2} \sim \mathcal P_{ij}  $& \\ \hline
		$  M_{ijk,0} \sim \mathcal Q^+_{3,ijk} $ & $ M_{ijk,1}\sim \mathcal Q^+_{2,ijk}  $&  $M_{ijk,2} \sim \mathcal K_{ijk}  $&  $M_{ijk,3} \sim \mathcal P_{ijk}  $ \\ \hline
	\end{tabular}
	\caption{First sets ($\ell \leq 3$) of parity-even conserved charges of non-radiative asymptotically flat spacetimes. The symbol $\sim$ means ``is proportional to''.}
	\label{tab:1}
\end{table}
\begin{table}[!htb]
	\centering
	\begin{tabular}{|c|c|c|c|}\hline
		$ S_{i,0} \sim \mathcal J_i  $& $ S_{i,1} \sim {\mathcal P}^-_i $& & \\   \hline 
		$ S_{ij,0} \sim \mathcal Q^-_{2,ij}  $& $ S_{ij,1} \sim \mathcal J_{ij}  $& $ S_{ij,2} \sim{\mathcal P}^-_{ij}  $& \\ \hline
		$  S_{ijk,0} \sim \mathcal Q^-_{3,ijk} $ & $ S_{ijk,1}\sim \mathcal Q^-_{2,ijk}  $&  $S_{ijk,2} \sim \mathcal J_{ijk}  $&  $S_{ijk,3} \sim {\mathcal P}^-_{ijk} $ \\ \hline
	\end{tabular}

	\caption{First sets ($\ell \leq 3$) of parity-odd conserved charges of non-radiative asymptotically flat spacetimes. The symbol $\sim$ means ``is proportional to''.}
	\label{tab:2}
\end{table}

We can further express the non-radiative multipolar conserved charges $M^\pm_{L,k}$, $0\leq k\leq \ell$ in the language of \cite{Freidel:2021ytz} as follows. Let us first consider $0 \leq k \leq \ell-2$. In that range, using Eqs. \eqref{dictionary_charge} and \eqref{23} and substituting $n = \ell-k$ we obtain 
\begin{eqnarray}
   M^\pm_{L,k} &=& m_{\ell,k}\mathcal Q^{R\pm}_{\ell-k,L},\qquad m_{\ell,k} \equiv \frac{2^{\ell-k}(2\ell+1)!!}{(2\ell-k)!(\ell+2)(\ell+1)}. 
\end{eqnarray}
Using Eq. \eqref{QsR} to substitute $\mathcal Q^{R\pm}_{\ell-k,L}=2\,  \text{Re}\, Q_{\ell-k}(\tau_{-\ell+k,\pm ,L})=\mp 2\,  \text{Im}\, Q_{\ell-k}(\tau_{-\ell+k,\mp ,L})$ we obtain 
\begin{eqnarray}
\boxed{   M^\pm_{L,k} = 2 m_{\ell,k} \text{Re}\, Q_{\ell-k}(\tau_{-\ell+k,\pm,L}) },\label{rel1}
\end{eqnarray}
where we choose to write the right-hand side in terms of the real part of the charge. Remember from Tables \ref{tab:1} and \ref{tab:2} that the cases $k=\ell-1$ and $k=\ell$ complete the expressions of the non-radiative multipoles in terms of the celestial charges. In fact, equation \eqref{rel1} is also true for $k=\ell-1$ and $k=\ell$ as a consequence of Eqs. \eqref{lowpolariz}, \eqref{loworders}, \eqref{PLexpr}, \eqref{KLexpr}, \eqref{JLexpr} and \eqref{ML}. This is a non-trivial check of our normalization factors for $n=0,1$ since the generic $n \geq 2$ cases are now extended to $n=0,1$ with the single numerical factor $m_{\ell,\ell-n}$. The equality \eqref{rel1} therefore holds for the entire range $0 \leq k \leq \ell$. 

Note that gravitational electric-magnetic duality under a $\pi/2$ rotation maps $M_L^\pm $ as \eqref{actionML}. It therefore maps $M^+_{L,k} \mapsto M_{L,k}^-$ and $M^-_{L,k} \mapsto -M^+_{L,k}$, or, equivalently,
\begin{equation}
Q_{s}(\tau^{}_{-s,\pm,L}) \mapsto \pm Q_{s}(\tau^{}_{-s,\mp ,L}).
\end{equation}

Finally, the algebra of celestial multipoles $M^\pm_{L,k}$ can be deduced from the algebra of the real part of the helicity $s$ complex charges using the dictionary \eqref{rel1}. Using complex algebra, we can write explicitly 
\begin{align} 
    \{\text{Re}\, Q_s(\tau), \text{Re}\, Q_{s'}(\tau')\}^{\text{lin}}&=\frac{1}{2}\text{Re}\, \{ Q_s(\tau),  Q_{s'}(\tau')\}^{\text{lin}}+\frac{1}{2}\text{Re}\, \{ Q_s(\tau),   Q^*_{s'}(\tau')\}^{\text{lin}}. 
\end{align}
The first bracket $\{ Q_s(\tau), Q_{s'}(\tau')\}^{\text{lin}}$ is given in Eq.~\eqref{charge_algebra} as derived in \cite{Freidel:2021ytz}. The second bracket $\{ Q_s(\tau), Q^*_{s'}(\tau')\}^{\text{lin}}$ involving the complex conjugated charge $Q^*_{s'}(\tau')$ was not derived in \cite{Freidel:2021ytz}. Completing the evaluation of this second bracket is the last step in order to obtain the algebra of non-radiative multipole moments. We leave the remaining part of the  derivation of the algebra of non-radiative multipole moments for further work.

\section*{Acknowledgements}
We thank Luc Blanchet and Guillaume Faye for collaboration on closely related topics.
We acknowledge fruitful discussions with Miguel Campiglia, Laurent Freidel and Jakob Salzer. R.O. thanks Simone Speziale and A.S. thanks  Marc Geiller and Etera Livine for discussions on related topics. G.C. is Senior Research Associate of
the F.R.S.-FNRS and acknowledges support from the FNRS research credit J.0036.20F, bilateral Czech convention
PINT-Bilat-M/PGY R.M005.19 and the IISN convention 4.4503.15. 
The work of R.O. is supported by the R\'egion \^Ile-de-France within the DIM ACAV$^{+}$ project SYMONGRAV (Sym\'etries asymptotiques et ondes gravitationnelles). A.S. is partially supported by the
ERC Advanced Grant ``High-Spin-Grav", and the ERC Starting Grant ``Holography for realistic black holes''.

\bibliography{ref_multipoles}

\providecommand{\href}[2]{#2}\begingroup\raggedright\begin{thebibliography}{10}

\bibitem{Strominger:2021lvk}
A.~Strominger, {\it {w(1+infinity) and the Celestial Sphere}},
  \href{http://arXiv.org/abs/2105.14346}{{\tt 2105.14346}}.

\bibitem{Strominger:2021mtt}
A.~Strominger, {\it {$w_{1+\infty}$ Algebra and the Celestial Sphere: Infinite
  Towers of Soft Graviton, Photon, and Gluon Symmetries}},  Phys. Rev. Lett.
  {\bf 127} (2021), no.~22 221601.

\bibitem{Guevara:2021abz}
A.~Guevara, E.~Himwich, M.~Pate and A.~Strominger, {\it {Holographic symmetry
  algebras for gauge theory and gravity}},  JHEP {\bf 11} (2021) 152
  [\href{http://arXiv.org/abs/2103.03961}{{\tt 2103.03961}}].

\bibitem{Freidel:2021ytz}
L.~Freidel, D.~Pranzetti and A.-M. Raclariu, {\it {Higher spin dynamics in
  gravity and $w_{1 + \infty}$ celestial symmetries}},
  \href{http://arXiv.org/abs/2112.15573}{{\tt 2112.15573}}.

\bibitem{Bondi:1960jsa}
H.~Bondi, {\it {Gravitational Waves in General Relativity}},  Nature {\bf 186}
  (1960), no.~4724 535--535.

\bibitem{1962RSPSA.269...21B}
H.~{Bondi}, M.~G.~J. {van der Burg} and A.~W.~K. {Metzner}, {\it {Gravitational
  Waves in General Relativity. VII. Waves from Axi-Symmetric Isolated
  Systems}},  Proceedings of the Royal Society of London Series A {\bf 269}
  (Aug., 1962) 21--52.

\bibitem{1962PhRv..128.2851S}
R.~{Sachs}, {\it {Asymptotic Symmetries in Gravitational Theory}},  Physical
  Review {\bf 128} (Dec., 1962) 2851--2864.

\bibitem{10.2307/2415610}
M.~G.~J. van~der Burg, {\it Gravitational waves in general relativity. ix.
  conserved quantities},  Proceedings of the Royal Society of London. Series A,
  Mathematical and Physical Sciences {\bf 294} (1966), no.~1436 112--122.

\bibitem{1962RSPSA.270..103S}
R.~K. {Sachs}, {\it {Gravitational Waves in General Relativity. VIII. Waves in
  Asymptotically Flat Space-Time}},  Proceedings of the Royal Society of London
  Series A {\bf 270} (Oct., 1962) 103--126.

\bibitem{Barnich:2009se}
G.~Barnich and C.~Troessaert, {\it {Symmetries of asymptotically flat 4
  dimensional spacetimes at null infinity revisited}},  Phys. Rev. Lett. {\bf
  105} (2010) 111103 [\href{http://arXiv.org/abs/0909.2617}{{\tt 0909.2617}}].

\bibitem{Barnich:2010eb}
G.~Barnich and C.~Troessaert, {\it {Aspects of the BMS/CFT correspondence}},
  JHEP {\bf 05} (2010) 062 [\href{http://arXiv.org/abs/1001.1541}{{\tt
  1001.1541}}].

\bibitem{Barnich:2011mi}
G.~Barnich and C.~Troessaert, {\it {BMS charge algebra}},  JHEP {\bf 12} (2011)
  105 [\href{http://arXiv.org/abs/1106.0213}{{\tt 1106.0213}}].

\bibitem{Kapec:2014opa}
D.~Kapec, V.~Lysov, S.~Pasterski and A.~Strominger, {\it {Semiclassical
  Virasoro symmetry of the quantum gravity $ \mathcal{S}$-matrix}},  JHEP {\bf
  08} (2014) 058 [\href{http://arXiv.org/abs/1406.3312}{{\tt 1406.3312}}].

\bibitem{Campiglia:2014yka}
M.~Campiglia and A.~Laddha, {\it {Asymptotic symmetries and subleading soft
  graviton theorem}},  Phys. Rev. {\bf D90} (2014), no.~12 124028
  [\href{http://arXiv.org/abs/1408.2228}{{\tt 1408.2228}}].

\bibitem{Campiglia:2020qvc}
M.~Campiglia and J.~Peraza, {\it {Generalized BMS charge algebra}},  Phys. Rev.
  D {\bf 101} (2020), no.~10 104039
  [\href{http://arXiv.org/abs/2002.06691}{{\tt 2002.06691}}].

\bibitem{Compere:2018ylh}
G.~Comp\`ere, A.~Fiorucci and R.~Ruzziconi, {\it {Superboost transitions,
  refraction memory and super-Lorentz charge algebra}},  JHEP {\bf 11} (2018)
  200 [\href{http://arXiv.org/abs/1810.00377}{{\tt 1810.00377}}]. [Erratum:
  JHEP 04, 172 (2020)].

\bibitem{Godazgar:2018vmm}
H.~Godazgar, M.~Godazgar and C.~N. Pope, {\it {Subleading BMS charges and fake
  news near null infinity}},  JHEP {\bf 01} (2019) 143
  [\href{http://arXiv.org/abs/1809.09076}{{\tt 1809.09076}}].

\bibitem{Godazgar:2018qpq}
H.~Godazgar, M.~Godazgar and C.~N. Pope, {\it {New dual gravitational
  charges}},  Phys. Rev. D {\bf 99} (2019), no.~2 024013
  [\href{http://arXiv.org/abs/1812.01641}{{\tt 1812.01641}}].

\bibitem{Godazgar:2018dvh}
H.~Godazgar, M.~Godazgar and C.~N. Pope, {\it {Tower of subleading dual BMS
  charges}},  JHEP {\bf 03} (2019) 057
  [\href{http://arXiv.org/abs/1812.06935}{{\tt 1812.06935}}].

\bibitem{Oliveri:2020xls}
R.~Oliveri and S.~Speziale, {\it {A note on dual gravitational charges}},  JHEP
  {\bf 12} (2020) 079 [\href{http://arXiv.org/abs/2010.01111}{{\tt
  2010.01111}}].

\bibitem{Freidel:2021fxf}
L.~Freidel, R.~Oliveri, D.~Pranzetti and S.~Speziale, {\it {The Weyl BMS group
  and Einstein\textquoteright{}s equations}},  JHEP {\bf 07} (2021) 170
  [\href{http://arXiv.org/abs/2104.05793}{{\tt 2104.05793}}].

\bibitem{Freidel:2021qpz}
L.~Freidel and D.~Pranzetti, {\it {Gravity from symmetry: duality and impulsive
  waves}},  JHEP {\bf 04} (2022) 125
  [\href{http://arXiv.org/abs/2109.06342}{{\tt 2109.06342}}].

\bibitem{Freidel:2021dfs}
L.~Freidel, D.~Pranzetti and A.-M. Raclariu, {\it {Sub-subleading Soft Graviton
  Theorem from Asymptotic Einstein's Equations}},
  \href{http://arXiv.org/abs/2111.15607}{{\tt 2111.15607}}.

\bibitem{Tamburino:1966zz}
L.~A. Tamburino and J.~H. Winicour, {\it {Gravitational Fields in Finite and
  Conformal Bondi Frames}},  Phys. Rev. {\bf 150} (1966) 1039--1053.

\bibitem{1985FoPh...15..605W}
J.~{Winicour}, {\it {Logarithmic asymptotic flatness}},  Foundations of Physics
  {\bf 15} (May, 1985) 605--616.

\bibitem{Grant:2021hga}
A.~M. Grant and D.~A. Nichols, {\it {Persistent gravitational wave observables:
  Curve deviation in asymptotically flat spacetimes}},  Phys. Rev. D {\bf 105}
  (2022), no.~2 024056 [\href{http://arXiv.org/abs/2109.03832}{{\tt
  2109.03832}}].

\bibitem{Blanchet:1985sp}
L.~Blanchet and T.~Damour, {\it {Radiative gravitational fields in general
  relativity I. general structure of the field outside the source}},  Phil.
  Trans. Roy. Soc. Lond. {\bf A320} (1986) 379--430.

\bibitem{Blanchet:1986dk}
L.~Blanchet, {\it {Radiative gravitational fields in general relativity. 2.
  Asymptotic behaviour at future null infinity}},  Proc. Roy. Soc. Lond. {\bf
  A409} (1987) 383--399.

\bibitem{Blanchet:1987wq}
L.~Blanchet and T.~Damour, {\it {Tail Transported Temporal Correlations in the
  Dynamics of a Gravitating System}},  Phys. Rev. {\bf D37} (1988) 1410.

\bibitem{Blanchet:1992br}
L.~Blanchet and T.~Damour, {\it {Hereditary effects in gravitational
  radiation}},  Phys. Rev. {\bf D46} (1992) 4304--4319.

\bibitem{Blanchet:2020ngx}
L.~Blanchet, G.~Comp\`ere, G.~Faye, R.~Oliveri and A.~Seraj, {\it {Multipole
  expansion of gravitational waves: from harmonic to Bondi coordinates}},  JHEP
  {\bf 02} (2021) 029 [\href{http://arXiv.org/abs/2011.10000}{{\tt
  2011.10000}}].

\bibitem{Bieri:2013ada}
L.~Bieri and D.~Garfinkle, {\it {Perturbative and gauge invariant treatment of
  gravitational wave memory}},  Phys. Rev. {\bf D89} (2014), no.~8 084039
  [\href{http://arXiv.org/abs/1312.6871}{{\tt 1312.6871}}].

\bibitem{Pasterski:2015tva}
S.~Pasterski, A.~Strominger and A.~Zhiboedov, {\it {New Gravitational
  Memories}},  JHEP {\bf 12} (2016) 053
  [\href{http://arXiv.org/abs/1502.06120}{{\tt 1502.06120}}].

\bibitem{Flanagan:2015pxa}
E.~E. Flanagan and D.~A. Nichols, {\it {Conserved charges of the extended
  Bondi-Metzner-Sachs algebra}},  Phys. Rev. D {\bf 95} (2017), no.~4 044002
  [\href{http://arXiv.org/abs/1510.03386}{{\tt 1510.03386}}].

\bibitem{Nichols:2018qac}
D.~A. Nichols, {\it {Center-of-mass angular momentum and memory effect in
  asymptotically flat spacetimes}},  Phys. Rev. D {\bf 98} (2018), no.~6 064032
  [\href{http://arXiv.org/abs/1807.08767}{{\tt 1807.08767}}].

\bibitem{Mitman:2020pbt}
K.~Mitman, J.~Moxon, M.~A. Scheel, S.~A. Teukolsky, N.~Deppe, L.~E. Kidder and
  W.~Throwe, {\it {Computation of Normal and Spin Memory in Numerical
  Relativity}},  \href{http://arXiv.org/abs/2007.11562}{{\tt 2007.11562}}.

\bibitem{Mitman:2021xkq}
K.~Mitman {\em et.~al.}, {\it {Fixing the BMS frame of numerical relativity
  waveforms}},  Phys. Rev. D {\bf 104} (2021), no.~2 024051
  [\href{http://arXiv.org/abs/2105.02300}{{\tt 2105.02300}}].

\bibitem{1964PhRv..136.1224P}
P.~C. {Peters}, {\it {Gravitational Radiation and the Motion of Two Point
  Masses}},  Physical Review {\bf 136} (Nov, 1964) 1224--1232.

\bibitem{PhysRev.131.435}
P.~C. Peters and J.~Mathews, {\it Gravitational radiation from point masses in
  a keplerian orbit},  Phys. Rev. {\bf 131} (Jul, 1963) 435--440.

\bibitem{Damour:1983tz}
T.~Damour, {\it Gravitational radiation reaction in the binary pulsar and the
  quadrupole formula controversy},  Phys. Rev. Lett. {\bf 51} (1983)
  1019--1021.

\bibitem{1983MNRAS.203.1049F}
M.~J. {Fitchett}, {\it {The influence of gravitational wave momentum losses on
  the centre of mass motion of a Newtonian binay system.}},  "Mon. Not. Roy.
  Astron. Soc." {\bf 203} (June, 1983) 1049--1062.

\bibitem{Blanchet:2013haa}
L.~Blanchet, {\it {Gravitational Radiation from Post-Newtonian Sources and
  Inspiralling Compact Binaries}},  Living Rev. Rel. {\bf 17} (2014) 2
  [\href{http://arXiv.org/abs/1310.1528}{{\tt 1310.1528}}].

\bibitem{Buonanno:2014aza}
A.~Buonanno and B.~S. Sathyaprakash, {\em {Sources of Gravitational Waves:
  Theory and Observations}}.
\newblock 10, 2014.
\newblock \href{http://arXiv.org/abs/1410.7832}{{\tt 1410.7832}}.

\bibitem{Blanchet:2018yqa}
L.~Blanchet and G.~Faye, {\it {Flux-balance equations for linear momentum and
  center-of-mass position of self-gravitating post-Newtonian systems}},  Class.
  Quant. Grav. {\bf 36} (2019), no.~8 085003
  [\href{http://arXiv.org/abs/1811.08966}{{\tt 1811.08966}}].

\bibitem{Compere:2019gft}
G.~Comp\`ere, R.~Oliveri and A.~Seraj, {\it {The Poincar\'e and BMS
  flux-balance laws with application to binary systems}},  JHEP {\bf 10} (2020)
  116 [\href{http://arXiv.org/abs/1912.03164}{{\tt 1912.03164}}].

\bibitem{Henneaux:2004jw}
M.~Henneaux and C.~Teitelboim, {\it {Duality in linearized gravity}},  Phys.
  Rev. D {\bf 71} (2005) 024018 [\href{http://arXiv.org/abs/gr-qc/0408101}{{\tt
  gr-qc/0408101}}].

\bibitem{Geroch:1970cd}
R.~P. Geroch, {\it {Multipole moments. II. Curved space}},  J. Math. Phys. {\bf
  11} (1970) 2580--2588.

\bibitem{Hansen:1974zz}
R.~O. Hansen, {\it {Multipole moments of stationary space-times}},  J. Math.
  Phys. {\bf 15} (1974) 46--52.

\bibitem{Nichols:2017rqr}
D.~A. Nichols, {\it {Spin memory effect for compact binaries in the
  post-Newtonian approximation}},  Phys. Rev. {\bf D95} (2017), no.~8 084048
  [\href{http://arXiv.org/abs/1702.03300}{{\tt 1702.03300}}].

\bibitem{Flanagan:2018yzh}
E.~E. Flanagan, A.~M. Grant, A.~I. Harte and D.~A. Nichols, {\it {Persistent
  gravitational wave observables: general framework}},  Phys. Rev. {\bf D99}
  (2019), no.~8 084044 [\href{http://arXiv.org/abs/1901.00021}{{\tt
  1901.00021}}].

\bibitem{Compere:2019odm}
G.~Comp\`ere, {\it {Infinite towers of supertranslation and superrotation
  memories}},  Phys. Rev. Lett. {\bf 123} (2019), no.~2 021101
  [\href{http://arXiv.org/abs/1904.00280}{{\tt 1904.00280}}].

\bibitem{Seraj:2021qja}
A.~Seraj, {\it {Gravitational breathing memory and dual symmetries}},  JHEP
  {\bf 05} (2021) 283 [\href{http://arXiv.org/abs/2103.12185}{{\tt
  2103.12185}}].

\bibitem{Seraj:2021rxd}
A.~Seraj and B.~Oblak, {\it {Gyroscopic Gravitational Memory}},
  \href{http://arXiv.org/abs/2112.04535}{{\tt 2112.04535}}.

\bibitem{Seraj:2022qyt}
A.~Seraj and B.~Oblak, {\it {The Precession Caused by Gravitational Waves}},
  \href{http://arXiv.org/abs/2203.16216}{{\tt 2203.16216}}.

\bibitem{1968RSPSA.305..175N}
E.~T. {Newman} and R.~{Penrose}, {\it {New Conservation Laws for Zero Rest-Mass
  Fields in Asymptotically Flat Space-Time}},  Proceedings of the Royal Society
  of London Series A {\bf 305} (June, 1968) 175--204.

\bibitem{Weinberg:1965nx}
S.~Weinberg, {\it {Infrared photons and gravitons}},  Phys. Rev. {\bf 140}
  (1965) B516--B524.

\bibitem{Cachazo:2014fwa}
F.~Cachazo and A.~Strominger, {\it {Evidence for a New Soft Graviton Theorem}},
   \href{http://arXiv.org/abs/1404.4091}{{\tt 1404.4091}}.

\bibitem{Hamada:2018vrw}
Y.~Hamada and G.~Shiu, {\it {Infinite Set of Soft Theorems in Gauge-Gravity
  Theories as Ward-Takahashi Identities}},  Phys. Rev. Lett. {\bf 120} (2018),
  no.~20 201601 [\href{http://arXiv.org/abs/1801.05528}{{\tt 1801.05528}}].

\bibitem{Godazgar:2019dkh}
H.~Godazgar, M.~Godazgar and C.~N. Pope, {\it {Dual gravitational charges and
  soft theorems}},  JHEP {\bf 10} (2019) 123
  [\href{http://arXiv.org/abs/1908.01164}{{\tt 1908.01164}}].

\bibitem{Campiglia:2016hvg}
M.~Campiglia and A.~Laddha, {\it {Subleading soft photons and large gauge
  transformations}},  JHEP {\bf 11} (2016) 012
  [\href{http://arXiv.org/abs/1605.09677}{{\tt 1605.09677}}].

\bibitem{Compere:2017wrj}
G.~Comp\`ere, R.~Oliveri and A.~Seraj, {\it {Gravitational multipole moments
  from Noether charges}},  JHEP {\bf 05} (2018) 054
  [\href{http://arXiv.org/abs/1711.08806}{{\tt 1711.08806}}].

\bibitem{Toth:2021cpx}
V.~T. Toth and S.~G. Turyshev, {\it {Efficient trace-free decomposition of
  symmetric tensors of arbitrary rank}},
  \href{http://arXiv.org/abs/2109.11743}{{\tt 2109.11743}}.

\bibitem{GHP}
R.~Geroch, A.~Held and R.~Penrose, {\it A spacetime calculus based on pairs of
  null directions},  Journal of Mathematical Physics {\bf 14} (1973), no.~7
  874--881 [\href{http://arXiv.org/abs/https://doi.org/10.1063/1.1666410}{{\tt
  https://doi.org/10.1063/1.1666410}}].

\bibitem{Ashtekar:1981bq}
A.~Ashtekar and M.~Streubel, {\it {Symplectic Geometry of Radiative Modes and
  Conserved Quantities at Null Infinity}},  Proc. Roy. Soc. Lond. {\bf A376}
  (1981) 585--607.

\bibitem{Ashtekar:1981sf}
A.~Ashtekar, {\it {Asymptotic Quantization of the Gravitational Field}},  Phys.
  Rev. Lett. {\bf 46} (1981) 573--576.

\bibitem{Campiglia:2015yka}
M.~Campiglia and A.~Laddha, {\it {New symmetries for the Gravitational
  S-matrix}},  JHEP {\bf 04} (2015) 076
  [\href{http://arXiv.org/abs/1502.02318}{{\tt 1502.02318}}].

\bibitem{Compere:2020lrt}
G.~Comp\`ere, A.~Fiorucci and R.~Ruzziconi, {\it {The $\Lambda$-BMS$_4$ charge
  algebra}},  JHEP {\bf 10} (2020) 205
  [\href{http://arXiv.org/abs/2004.10769}{{\tt 2004.10769}}].

\bibitem{Schoen:1979zz}
R.~Schoen and S.-T. Yau, {\it {Positivity of the Total Mass of a General
  Space-Time}},  Phys. Rev. Lett. {\bf 43} (1979) 1457--1459.

\bibitem{Ludvigsen:1981gf}
M.~Ludvigsen and J.~A.~G. Vickers, {\it {The Positivity of the Bondi Mass}},
  J. Phys. A {\bf 14} (1981) L389--L391.

\bibitem{Horowitz:1981uw}
G.~T. Horowitz and M.~J. Perry, {\it Gravitational energy cannot become
  negative},  Phys. Rev. Lett. {\bf 48} (1982) 371.

\bibitem{Thorne:1980ru}
K.~S. Thorne, {\it {Multipole Expansions of Gravitational Radiation}},  Rev.
  Mod. Phys. {\bf 52} (1980) 299--339.

\bibitem{1963JMP.....4..924M}
C.~W. {Misner}, {\it {The Flatter Regions of Newman, Unti, and Tamburino's
  Generalized Schwarzschild Space}},  Journal of Mathematical Physics {\bf 4}
  (July, 1963) 924--937.

\bibitem{Strominger:2013jfa}
A.~Strominger, {\it {On BMS Invariance of Gravitational Scattering}},  JHEP
  {\bf 07} (2014) 152 [\href{http://arXiv.org/abs/1312.2229}{{\tt 1312.2229}}].

\bibitem{He:2014laa}
T.~He, V.~Lysov, P.~Mitra and A.~Strominger, {\it {BMS supertranslations and
  Weinberg\textquoteright{}s soft graviton theorem}},  JHEP {\bf 05} (2015) 151
  [\href{http://arXiv.org/abs/1401.7026}{{\tt 1401.7026}}].

\bibitem{Choi:2019sjs}
S.~Choi and R.~Akhoury, {\it {Magnetic soft charges, dual supertranslations,
  and \textquoteright{}t Hooft line dressings}},  Phys. Rev. D {\bf 102}
  (2020), no.~2 025001 [\href{http://arXiv.org/abs/1912.02224}{{\tt
  1912.02224}}].

\bibitem{Satishchandran:2019pyc}
G.~Satishchandran and R.~M. Wald, {\it {Asymptotic behavior of massless fields
  and the memory effect}},  Phys. Rev. D {\bf 99} (2019), no.~8 084007
  [\href{http://arXiv.org/abs/1901.05942}{{\tt 1901.05942}}].

\end{thebibliography}\endgroup
\end{document}